\documentclass[showpacs,preprintnumbers,amsmath,amssymb,floatfix,
nofootinbib,aps,twocolumn]{revtex4}

\usepackage[dvips]{graphicx,graphics}
\usepackage{dcolumn}% Align table columns on decimal point
\usepackage{bm}
\usepackage{epsfig}
\usepackage{amsmath}
\newcommand{\la}{\langle}
\newcommand{\ra}{\rangle}

\newcommand{\beq}{\begin{eqnarray}}
\newcommand{\eeq}{\end{eqnarray}}
\newcommand{\sbeq}{\begin{subeqnarray}}
\newcommand{\seeq}{\end{subeqnarray}}

\newcommand{\bl}{\biggl}
\newcommand{\br}{\biggr}
\newcommand{\mbfr}{\mathbf{r}}

\newcommand{\bfk}{\mbox{{\boldmath $k$}}}

\newcommand{\bfv}{\mbox{{\boldmath $v$}}}
\newcommand{\bfr}{\mbox{{\boldmath $r$}}}
\newcommand{\bfj}{\mbox{{\boldmath $j$}}}

\newcommand{\bfq}{\mbox{{\boldmath $q$}}}

\newcommand{\bfJ}{\mbox{{\boldmath $J$}}}

\newcommand{\bftheta}{\mbox{{\boldmath $\theta$}}}

\newcommand{\bfV}{\mbox{{\boldmath $V$}}}
\begin{document}

\title{Dynamics near  QCD critical point \\ by dynamic renormalization group }
\author{Yuki Minami}\email{y-minami@ruby.scphys.kyoto-u.ac.jp}
\affiliation{Department of Physics, Kyoto University, Kyoto
606-8502, Japan}


\begin{abstract}
We work out the basic analysis on  dynamics near QCD critical point (CP) 
by dynamic renormalization group (RG).
In addition to the RG analysis by coarse graining,
we construct the nonlinear Langevin equation as a basic equation for the critical dynamics.
Our construction is based on the generalized Langevin theory 
and the relativistic hydrodynamics. 
Applying the dynamic RG to the constructed equation,
we derive the RG equation for the transport coefficients 
and analyze their critical behaviors.
We find that the resulting RG equation turns out to be the same 
as that for the liquid-gas CP except for an insignificant constant.
Therefore, the bulk viscosity and the thermal conductivity strongly diverge at the QCD CP.
We also show that the thermal and viscous diffusion modes exhibit critical slowing down
with the dynamic critical exponents $z_{\rm thermal}\sim 3$ and $z_{\rm viscous}\sim 2$ , respectively.
In contrast, the sound propagating mode shows critical speeding up
with the negative exponent $z_{\rm sound}\sim -0.8$. 
\end{abstract}

\maketitle

\section{Introduction}

An interesting feature of the phase diagram in  quantum chromodynamics (QCD)
is the possible existence of the critical point (CP), which 
is predicted by various effective models of QCD and suggested by
lattice QCD simulations.
The critical point is the end point of the first order phase transition line existing 
in the low temperature ($T$) region \cite{qcdcp}\footnote{
In fact,  the QCD matter seems to have
extremely rich structure in the phase 
diagram with one or multiple critical points \cite{qcdcp,Kitazawa:2002bc,Hatsuda:2006ps}, 
and even accompanied with inhomogeneous phases at low $T$ \cite{Alford:2000ze,Nakano:2004cd},
although the very existence of a CP may be questioned  according to
\cite{deForcrand:2006pv,Fukushima:2008is}.}.  
Then, the significance of this QCD CP is
that the phase transition at this point is of second order, and thereby 
we can expect critical phenomena due to large fluctuations of various quantities 
at  this point \cite{Stephanov:1998dy}.

Then, a fundamental problem arises;
what is the soft modes of the QCD CP?
A hint is that the baryon-number susceptibility \cite{Kunihiro:1991qu} diverges
at the CP as first suggested in \cite{Kunihiro:2000ap} and subsequently demonstrated in 
\cite{Hatta:2002sj}.
It has been established  that the fluctuating modes of conserved densities 
 are the soft modes at the CP \cite{fujii,son}.
Although the $\sigma$ mode seems to be the soft mode,
it  couples to the density fluctuation at finite density \cite{Kunihiro:1991qu} and
remains massive \cite{fujii,son,Stephanov:2008qz}\footnote{Such a fast mode as the $\sigma$ 
near the QCD CP is called a slaved mode \cite{haken}, because its slow dynamics is controlled
by  the density fluctuation for the QCD CP.}.

Furthermore, as a critical phenomenon,
some authors suggested a divergent behavior of bulk viscosity at  the QCD CP \cite{karsch},
although their validity of their argument is very much controversial
\cite{Romatschke:2009ng, moore, sasaki}; for instance,
 the ansatz  for the spectral function adopted in \cite{karsch}
may not  necessarily be true \cite{Romatschke:2009ng}, and 
a microscopic calculation by the relativistic Boltzmann equation \cite{sasaki} 
shows that the bulk viscosity is finite at the CP.
Thus, it is still not obvious whether 
 the transport coefficients will diverge at the QCD CP.

In fact, as is known in condensed matter physics, 
the critical divergence of the transport coefficients is a common phenomenon at a CP,
  such as  at the liquid-gas CP,
and originate from a universal mechanism; 
nonlinear fluctuations of {\em macroscopic} variables 
cause the divergence \cite{kawasaki1, mori:fujisaka}.
This implies that microscopic processes, as may be described by such as
the Boltzmann equation, would give only a minor contribution 
to the critical divergence of these quantities, if any.
The dynamic renormalization group (RG) theory \cite{onuki, mazenko}
is a  standard technique for critical dynamics,
which systematically incorporate the macroscopic fluctuations 
causing the  divergent behavior of transport coefficients.
In this theory, 
We must construct a nonlinear Langevin equation 
as a basic equation for the critical dynamics. 
The construction goes as follows.
First,  the slow variables are identified for describing the critical dynamics.
Next, the thermodynamic potential for the slow variables is constructed
to determine the static property of the system.  
Finally, the streaming
terms, causing the dynamic-nonlinear coupling,
 and the kinetic coefficients are determined
for respectively describing time reversible and irreversible changes of the slow variables.
We note that the streaming term is absent in the simple  Brownian motion.

The general theory of the critical dynamics as described above
tells us that an essential ingredient is 
to  properly construct the  nonlinear Langevin equation
for the critical dynamics.     
As far as we know, this is the first attempt for the QCD CP.
Our construction of the Langevin equation  
is based on the generalized Langevin theory, by Mori \cite{mori, mori:fujisaka}, and the {\em relativistic} hydrodynamics, 
because the slow variables are identified as long-wavelength fluctuations of the conserved densities 
\cite{fujii,son,Minami:2009hn};
we construct the streaming terms from continuity equations and 
the potential condition, which is a general condition for streaming terms \cite{onuki, mazenko}.
Also, we use the thermodynamic potential for 
the 3d Ising system as that for the QCD CP
because the static universality class is the same as 3d Ising class \cite{fujii,son,onukin}.
Finally, we determine the kinetic coefficients 
from a relativistic hydrodynamic equation, here  
the so called  Landau equation \cite{landau} taken.
In consequence, we shall show that 
the Langevin equation differs from it for the liquid-gas CP due to relativistic effects,
although the dynamic universality class of the QCD CP is conjectured as of the liquid-gas CP \cite{son, moore}.  

After such construction, 
we apply the dynamic RG to the Langevin equation and derive the RG equations for the transport coefficients.
Consequently, to our surprise,
these RG equations turn out to be the same as for the liquid-gas CP 
except for a irrelevant constant,
although the Langevin equations are different.    
Therefore, the bulk viscosity and the thermal conductivity strongly diverge 
and can be more important than the shear viscosity near the QCD CP.
We shall also show that the thermal and viscous diffusion modes exhibits critical slowing down,
whereas the sound mode critical speeding up.  

This paper is organized as follows. 
In Sec.II, we shortly review the theory of critical dynamics.
In Sec.III, we construct the nonlinear Langevin equation for the QCD CP
by the generalized Langevin theory and the relativistic hydrodynamics.
In Sec.IV,  we analyze the critical exponents of the transport coefficients and dynamic critical exponents
by the dynamic RG.
The final section is devoted to a summary and concluding remarks.
In Appendixes \ref{sec: derivation}, \ref{sec: RG1} and \ref{sec: RG2}, 
we give the detailed derivation of the RG equations for the transport coefficients.

\section{Review of critical dynamics}
Since the present work is based on the 
theory of critical dynamics in condensed matter physics \cite{onuki,mazenko},
we now shortly review it for self-containedness.

\subsection{Critical divergences of transport coefficients}

The critical divergence of transport coefficients (or diffusion constants) is
a common phenomenon, for instance, 
at the critical point of the liquid-gas, ferromagnetic transitions  and so on
\cite{onuki, mazenko}.
The important point is
that the critical divergence 
 originates from a universal mechanism; 
 nonlinear fluctuations of macroscopic variables 
causes the divergence \cite{kawasaki1, mori:fujisaka}.

Here, we briefly illustrate how macroscopic nonlinear fluctuations
 cause the critical divergence, taking 
the thermal conductivity near the liquid-gas CP as an example \cite{kawasaki2}.
The thermal conductivity is given by the Kubo formula as follows,
\begin{equation}
\lambda = T^{-2}\int d{\mbfr}\int^{\infty}_0 dt \la q(\mbfr,t) q(0,0) \ra,
\label{eq:heat-kubo}
\end{equation}
where $q(\mbfr,t)$ and $T$ are the heat current and temperature, respectively, 
and $\la \cdots \ra$ denotes the statistical average in the thermal equilibrium state.
The heat current $q(\mbfr,t)$  
is supplied from two sources:
one is due to a microscopic process as calculated by a microscopic theory, such as
the Boltzmann equation, and the other is  
due to  nonlinear fluctuations of macroscopic variables \cite{mori}:
\begin{equation}
q = q_{\rm micro}+q_{\rm macro},
\end{equation}
where $ q_{\rm micro}$ and $q_{\rm macro}$ respectively denote
 the microscopic and macroscopic currents.
The macroscopic process causing the heat current is identified as
the entropy density convected by
fluid velocity fluctuation. Thus,
we have
\begin{equation}
q_{\rm macro}\sim \delta s \delta v,
 \label{eq: hc}
\end{equation}
where $\delta s$ and $\delta v$
 respectively denote the fluctuations of  the entropy density and  the fluid velocity.
The macroscopic current Eq. (\ref{eq: hc}) is of the second order in fluctuations 
and hence negligible far from the CP.
However, it becomes the dominant part near the CP,
since the fluctuations are enhanced there.
We see that Eq. (\ref{eq:heat-kubo}) 
now has the following form
\begin{equation}
\lambda = \lambda_{\rm micro}+\int d\mbfr\int_0^{\infty} dt 
\la\delta s(\mbfr,t) \delta v(\mbfr,t)\delta s(0,0)\delta v(0,0) \ra,
\label{eq: tc}
\end{equation}
where $\lambda_{\rm micro}$ is the thermal conductivity coming from the microscopic current.
Recalling that the entropy density fluctuation is a soft mode near the liquid-gas CP,
we see that the second term of Eq. (\ref{eq: tc}) diverges at the CP.
This is the mechanism causing the critical divergence of transport coefficients.

Let us call the transport coefficients, such as $\lambda_{\rm micro}$, coming from  microscopic processes
the bare transport coefficients,
and those including the  contributions from the nonlinear macroscopic fluctuations 
the renormalized ones.
Then, we need not to study the critical divergence of  transport coefficients by a microscopic theory 
since the divergence originates from only the macroscopic processes.
The dynamic RG \cite{onuki, mazenko, siggia, kg, onukin} is the 
standard theory incorporating such nonlinear macroscopic fluctuations.
In this theory, we must construct a nonlinear Langevin equation 
as a basic equation for the critical dynamics. 

\subsection{Generalized Langevin equation}

We first note that if  the dynamic variables are
divided into slow and fast ones, the slow dynamics can be well described by a Langevin equation.
We stress that such the Langevin equation can be 
 derived in a generic way, so-called the Mori theory \cite{mori, mori:fujisaka},  
from the microscopic equation of motion.
The starting microscopic equations of motion 
are  Liouville or Hisenberg equation for a classical or a quantum system, respectively.
They read
\begin{equation}
\frac{\partial }{\partial t}A_j (t)=\{ A_{j}(t), H\}_{\rm PB}, \label{eq: liouville} 
\end{equation}
and 
\begin{equation}
\frac{\partial }{\partial t}A_j(t)=(1/i\hbar)[A_{j}(t), H], \label{eq: hisenberg}
\end{equation}
respectively. Here, 
$A_j(t) $ are arbitrary slowly varying variables (slow variables), 
$H$ a microscopic Hamiltonian, and, 
$\{\cdot,\cdot\}_{\rm PB}$ and $[\cdot,\cdot]$ represent
  Poisson brackets and commutation relations, respectively.
Equation (\ref{eq: liouville}) (or Eq. (\ref{eq: hisenberg})) can be divided into 
 two parts: one 
is composed of only the slow variables and describes their slow motion, 
while the other involves  fast motions
due to  the microscopic processes. Thus, we have \cite{mori:fujisaka, onuki} 
\begin{equation}
\frac{\partial}{\partial t}A_j(t)=v_j(A)
-\sum\limits_{k} L_{j k}(A) \frac{\delta (\beta H(A) )}{\delta A_k}+\theta_j (t),
\label{eq: nllangevin}
\end{equation}
with $\beta$ being the inverse temperature.
Here, 
$v_j(A(t))$, which is called a streaming term,
  gives the time-reversible
process, while
 $L_{j k}$ and $H(A(t))$ are
bare kinetic coefficients and 
a thermodynamic potential for $A_j$, respectively.
The first and second terms in Eq. (\ref{eq: nllangevin}) 
 are the slow motions and nonlinear in $A_j$, whereas
the last term is the fast motions
and  treated as a stochastic variable
obeying 
the fluctuation-dissipation relation
\begin{equation}
\la \theta_j (t) \theta_k(t^{'}) ;a \ra = 2 L_{j k}(a)\delta(t-t^{'}). \label{eq: fdr}
\end{equation}
Here, $\la \cdots ; a\ra$ represents the conditional average in which $A_j$ is fixed at $a_j$.
We stress that this relation is not given by hand, 
but naturally obtained from the decomposition process \cite{mori:fujisaka}.   

Equation (\ref{eq: nllangevin}) is called the generalized Langevin equation,
which has been widely used in the phase transition dynamics \cite{onuki, mazenko}.
Even for the QCD CP, we may use the generalized Langevin equation, because 
 only the time-scale separation is assumed in the Mori theory.
Furthermore, we note that, by the time-scale separation, transport coefficients arises .

\subsection{Dynamic RG}

The general dynamic RG transformation usually 
consists of  two procedures, 
i.e., coarse graining and rescaling as in the
static RG transformation \cite{hhs, mazenko}.
However, as is shown in \cite{kg, onukin, onuki},
we can omit the rescaling, 
if we are interested in only  the critical exponents of transport coefficients,
although the relevant fixed point seems to be absent
in such a simplified RG transformation \cite{kg}.

The Langevin equation is an infrared effective theory
and  inherently has a ultraviolet cutoff $\Lambda_0$, 
which should satisfy the following inequality
\begin{equation} 
\xi^{-1} \ll \Lambda_0 \ll a^{-1}.
\end{equation}
Here, $\xi$ and $a$ are the correlation length and the characteristic microscopic length,
respectively.
Then, the  Langevin equation is coarse grained by
averaging over the large-wavenumber components
 of the slow variables $A_j(t)$ in the infinitesimal wavenumber shell,
\begin{equation} 
\Lambda-\delta \Lambda < k <\Lambda,
\end{equation}
for Eq. (\ref{eq: nllangevin}).
Here, $\Lambda$ starts from the initial value $\Lambda_0$ and is lowered up to 
$\Lambda \ll \xi^{-1}$.   
In this way, we infinitesimally make coarse graining of the Langevin equation. 
Because the coarse-graining procedure is infinitesimal, we need not the rescaling .
Inspecting the form of the coarse-grained Langevin equation, 
we can obtain the RG equation for the transport coefficients.

\subsection{Contrast with the static RG }

Here, we first stress that 
the concept of the dynamic universality class is not so universal contrary to its name.
Then, the class of the QCD CP 
may not be the same as of the liquid-gas CP or the model H
\footnote{The model H \cite{kawasaki1, HALPERIN:1969zza} is the minimal-dynamic model for a CP
that its relevant modes are given as the nonrelativistic-hydrodynamic modes. 
The liquid-gas CP belongs to the dynamic universality class of the model H}, 
although it is conjectured by \cite{son, moore}.
To see this, let us contrast the difference
 between the static RG with the dynamic one.

An important point is that the respective infrared effective theories are different;
in the static case, 
the infrared effective theory is the thermodynamic potential (or so-called Landau free energy), 
the nature of which turns out to be governed by only
the space dimension and the symmetry among the order parameters but
not by the details of the dynamics, and hence the concept of the universality class makes sense for the static case.
In contrast, for the dynamic RG,
the infrared effective theory is the nonlinear Langevin equation.

Here, the important difference arises; 
 the relevant variables for the Langevin equation is not only the order parameters 
but also conserved densities, and its nonlinear couplings can not be determined by only the symmetry in general.
Consequently,  the dynamic universality class is not so universal compared to the static one.
Specifically, the nonlinear couplings, namely, the streaming terms $v_j(A)$, 
are generally given by the Poisson brackets (commutation relations)
 among the slow variables in the classical (quantum) system \cite{mazenko};
\begin{eqnarray}
v_j(A)=\sum\limits_{k}[Q_{j k}(A)\frac{\delta H}{\delta A_k}
-\beta^{-1}\frac{\delta }{\delta A_k}Q_{j k}(A)] ,
\label{eq: stream}
\end{eqnarray}
where
\begin{equation}
Q_{j k}(A)= \la \{A_j, A_k \}_{\rm PB} ; A \ra \; \mbox{or}\;  \la [ A_j, A_k ]/(i \hbar)  ; A\ra. 
\end{equation} 
The important point is that the Poisson-bracket relations depend on the microscopic expressions of the variables. 
This fact leads to an important consequence that the dynamic universality class of the QCD CP 
may not be the same as  of the liquid-gas CP or the model H.
Actually, in the model H, the Poisson-bracket relations are calculated with
 the non-relativistic relations \cite{kawasaki1, mazenko}.

\section{The nonlinear Langevin equation for the QCD CP}

\subsection{Slow variables}

We first identify the slow variables near the QCD CP,
which consist of  soft modes and conserved densities.
For the QCD CP, the soft modes are nothing but the long-wavelength fluctuations 
of the conserved densities, i.e., 
the baryon number density $n$ and  the energy and momentum $T_{\mu \nu}$ \cite{fujii, son}.
Thus, we see that the slow variables near the QCD CP are 
given by only the fluctuations of the conserved quantities:
\begin{equation} 
A_j =\{ \delta n, \delta e=(\delta T^{0 0}), \delta J^i=(\delta T^{0 i}) \}.
\end{equation}
Because the slow dynamics of the conserved quantities 
is basically given by hydrodynamics,
we find that the system near the QCD CP is described as a relativistic critical fluid.
In other words, the relevant modes are given as the relativistic-hydrodynamic modes.
This is the basic observation for our construction of 
 the nonlinear Langevin equation for the QCD CP.
More specifically,
the hydrodynamic modes are the 
 thermal and viscous diffusion modes, and the sound propagating mode.
The thermal mode is the entropy fluctuation inducing 
 the density and energy fluctuations,
whereas the viscous and the sound modes are the transverse and longitudinal components
of the momentum fluctuations, respectively. 

Now, we note that not all fluctuations are enhanced near the CP.
Therefore, we can neglect nonlinearity of fluctuations that is not enhanced,
if such fluctuations are identified.
Then, let us identify the non-enhanced fluctuations by the hydrodynamics.
The usual hydrodynamics with static scaling laws
is useful to see the such tendency of the slow variables.
Since the result turns out to be independent of the choice of the frame, 
which defines the local rest frame \cite{Minami:2009hn},
let us take the energy frame, namely the Landau equation \cite{landau}, 
which is given by the following conservation laws:
\begin{eqnarray}
\partial_{\mu}N^{\mu}=0, \label{eq: n}\\
\partial_{\mu}T^{\mu\nu}=0, 
\end{eqnarray}
where $N^{\nu}$ and $T^{\mu \nu}$ are the particle current and the energy-momentum tensor, respectively.
Those are given as
\begin{eqnarray}
N^\mu &=& n u^\mu+\nu^\mu, \\
T^{\mu \nu}&=&h u^{\mu}u^{\nu}-Pg^{\mu\nu}+\tau^{\mu\nu},  
\end{eqnarray}
where $h=e+P$ is the enthalpy density with $e$ 
and $P$ being the energy density and the pressure.
Also, $u^{\mu}=(\gamma, \gamma \bfv)$ are the fluid four velocity,
with $\gamma$ being the Lorentz factor, and
the dissipative terms, $\nu^{\mu}$ and $\tau^{\mu  \nu}$, are given by 
\begin{eqnarray}
\nu^{\mu} =&& \lambda_0 \bl( \frac{n T}{h} \br)^2 \partial_{\perp}^\mu \bl(\frac{\mu}{T} \br), \label{eq: d1} \\
\tau^{\mu \nu}=&&\eta_0 [\partial^{\mu}_{\perp}u^{\nu}
               +\partial^{\nu}_{\perp}u^{\mu}
               -\frac{2}{3}\Delta^{\mu\nu}(\partial_{\perp}{\cdot}u)] \nonumber \\
               &&+\zeta_0\Delta^{\mu\nu}(\partial_{\perp}{\cdot}u), \label{eq: d2}                         
\end{eqnarray} 
where $\lambda_0$, $\eta_0$ and $\zeta_0$ are the bare thermal conductivity, the bare share and 
 bulk viscosities, respectively.
 $\Delta^{\mu\nu} \equiv g^{\mu \nu}-u^{\mu} u^{\nu}$ is the projection onto the space-like vector 
and $\partial^{\mu}_{\perp} \equiv \Delta^{\mu \nu}\partial_{\nu}$ is  the space-like derivative.

In the hydrodynamic regime, $k\xi \ll 1$,
 the hydrodynamic modes is analyzed by the linearized equation, which is given by
\begin{eqnarray}
\frac{\partial \delta n}{\partial t}=&&-n_c \nabla \cdot \delta \bfv 
                                     +\lambda_0 \bl( \frac{n_c T_c}{h_c} \br)^2 \nabla^2 \delta 
                                        \bl( \frac{\mu}{T}\br),
   \label{eq: ln}\\
\frac{\partial \delta e}{\partial t}=&& - h_c\nabla \cdot \delta \bfv , \label{eq: le}\\
\frac{\partial \delta\bfJ}{\partial t}=&& -\nabla(\delta P)
  +(\zeta_0+\frac{1}{3}\eta_0)\nabla(\nabla\cdot\delta\bfv ) \\ \nonumber
   &&+\eta_0\nabla^{2}\delta\bfv ,   \label{eq: lj}
\end{eqnarray}
where the symbols with a prefix $\delta$
 denote the fluctuations from their equilibrium values, which are
 denoted by a suffix $c$
\footnote{ Here, we have slightly rewritten 
the form of the linearized equations, (\ref{eq: ln})- (\ref{eq: lj}), 
 from those in \cite{Minami:2009hn} by the thermodynamic relations
$\delta \epsilon=T\delta(ns)+ \mu \delta n$ and $\delta P= ns \delta T + n\delta \mu$,
where $s$ is the entropy per particle. }.
Hereafter, variables with the suffix and the prefix respectively denote the
equilibrium values and fluctuations.   
As relativistic effects, we see that dissipative effects appear in Eq. (\ref{eq: ln}) 
while vanish in Eq. (\ref{eq: le}),
because we have chosen  the energy frame.
We note that the relativistic effect for the particle frame appears in a different form \cite{Minami:2009hn}.    
   
By Eqs. (\ref{eq: ln})-(\ref{eq: lj}) and static scaling laws,
 the tendency of the hydrodynamic modes is analyzed in the critical regime, $k\xi \gg 1$.
We have studied the such behavior in the previous paper \cite{Minami:2009hn}
and shown that the thermal mode is enhanced 
, whereas the sound mode is suppressed 
and the viscous mode is not enhanced nor suppressed.
Recalling the relation between the hydrodynamic modes and the slow variables, 
we have the result that $\delta n$ 
and  $\delta e$ are enhanced, 
while $\delta \bfJ$ is not near the QCD CP.

We note that 
nonlinear couplings among these modes, which is not included in usual hydrodynamics,
become significant in the critical regime.  
We will take them into account in the nonlinear Langevin equation,
except for the fluctuation of  the momenta $\delta \bfJ$, 
 the nonlinear term of which will be neglected even in the critical regime.

\subsection{Thermodynamic potential for the slow variables}

Next, we construct the thermodynamic potential 
$H(\delta n,\delta e, \delta \bfJ)$ for the slow variables.

Since the momentum density fluctuation is not enhanced near the QCD CP,
we can neglect its coupling with $\delta n$ and $\delta e$, and  
may adopt a Gaussian form for the momentum density part of the potential.
Thus, we have
$H(\delta n,\delta e, \bfJ)=H_{n e}(\delta n,\delta e)+H_{J}(\delta \bfJ)$,
with
\begin{equation}
H_{J}(\delta \bfJ) \equiv \frac{1}{2 h_c}\delta \bfJ^2.
\end{equation} 
In contrast to $\delta \bfJ$,
$\delta n$ and $\delta e$
are significantly enhanced near the QCD CP, the thermodynamic
potential $H_{ n e}(\delta n,\delta e)$ should contain 
 higher order terms of them.
In fact,  $H_{ n e}(\delta n, \delta e)$ is 
the quantity that determines the static property of the system
and the QCD CP belongs to the same static universality class as the 3d Ising class, namely, $Z_2$.
Therefore, we may construct $H_{n e}(\delta n, \delta e)$ 
with the thermodynamic potential for the 3d Ising system \cite{glw}, 
which reads
\begin{eqnarray}
\beta H_{\rm Ising}(\psi, m) =&&\int d\bfr 
   [\frac{1}{2}r_0\psi^2+\frac{1}{2}K_0|\nabla \psi|^2
   +\frac{1}{4}u_0\psi^4  \nonumber \\ &&
   +\gamma_0 \psi^2 m +\frac{1}{2 C_0}m^2-h\psi-\tau m ] .\label{eq: glw}
\end{eqnarray}
Here, $\psi$  and $m$ are the spin density and the exchange energy density, respectively.
$r_0$, $K_0$, $u_0$, $\gamma_0$ and $C_0$ denote the static parameters,
while $h$ and $\tau$ 
 the applied magnetic field and the reduced temperature,
respectively.
Then, we assume that  the thermodynamic potential  takes the following form
\begin{equation}
H(\delta n,\delta e, \delta \bfJ )=H_{\rm Ising}(\psi, m)
 + \frac{1}{2 h_c}\delta \bfJ^2, \label{eq: potential} 
\end{equation}
provided that the mapping between $(\psi,\, m)$ and
$(\delta n,\, \delta e)$ is given.

The general mapping relation between a grand canonical ensemble in $Z_2$ and the 3d Ising system is
known in condensed matter physics \cite{onuki}, which are summarized as follows.
First,
we assume the following  linear relation between the deviations of the respective intensive variables
from those at the critical points, which should be valid near the CP.
\footnote{Recall that the static scaling laws
are expressed by the deviations of the intensive variables from those at the CP.}:
\begin{eqnarray}
\delta h=&&\alpha_1\delta (\mu/T) +\alpha_2 \delta T/T_c , \label{eq: mr1}\\
\delta \tau =&&\beta_1\delta (\mu/T) +\beta_2 \delta T/T_c, \label{eq: mr2}
\end{eqnarray}
where 
$\alpha_1$, $\alpha_2$, $\beta_1$ and $\beta_2$ are constants and
assumed to be regular at the CP.
We note that $\alpha_1$, $\alpha_2$, $\beta_1$ and $\beta_2$ need not to be determined for
 the critical divergence of transport coefficients,
since they have no singularities at the CP. 
Although one could use 
Eqs. (\ref{eq: mr1}) and (\ref{eq: mr2}) for the mapping,
  it turns out to be inconvenient 
for the analysis by a Langevin equation.
To translate these equations to more convenient ones,
we assume the following relation \cite{onuki}:
\begin{equation}
\psi\delta h  + m \delta \tau 
=T_c^{-2}\delta T\delta e+\delta(\mu/T)\delta n, \label{eq: mr3}
\end{equation}
which is actually derived by considering 
a change of the microscopic distributions 
due to small deviations of the external parameters in both systems.
From the relations Eqs. (\ref{eq: mr1})-(\ref{eq: mr3}), we arrive at the
convenient mapping relation as follows,
\begin{eqnarray}
\delta n =&& \alpha_1\psi +\beta_1 m,
\label{eq: map1} \\
T_c^{-1}\delta e =&& \alpha_2\psi +\beta_2 m. 
\label{eq: map2}
\end{eqnarray}
With this mapping,
Eq. (\ref{eq: potential}) now gives the thermodynamic potential for the QCD CP.
We remark that we only map the static quantities
 although the dynamic ones are studied. 
For later uses, we introduce fluctuations of the intensive variables as
\begin{eqnarray}
\delta T &&\equiv T_c^2 \frac{\delta(\beta H)}{\delta e}, \label{eq: tf}\\ 
\delta \bl( \frac{\mu}{T} \br) &&\equiv \frac{\delta(\beta H)}{\delta n} .\label{eq: muf}
\end{eqnarray}
This relation comes from the fact that, 
in the grand canonical distribution $P_{\rm gra} \propto \exp[(1/T) e + (\mu/T) n]$,
 $e$ and $n$ are respectively conjugate to $1/ T$ and $\mu /T$
\cite{onuki}.
We also introduce the fluid velocity fluctuation as in the non-relativistic case:
\begin{equation}
\delta \bfv \equiv  \frac{\delta H}{\delta \bfJ}. \label{eq: vf}
\end{equation}

We note that the static parameters in Eq. (\ref{eq: glw}) 
has the ultraviolet cutoff dependence in the region $\xi^{-1}<\Lambda$.
Let us write the static parameters as 
$r(\Lambda)$,  $K(\Lambda)$, $u(\Lambda)$,  $\gamma(\Lambda)$ and $C(\Lambda)$ to make
 their $\Lambda$ dependence explicit.
These variables have the following asymptotic behaviors \cite{onukin,onuki,glw}:
\begin{eqnarray}
r(\Lambda ) &&\sim\Lambda^{2-\eta}, \label{eq: rlambda}\\
K(\Lambda ) &&\sim\Lambda^{-\eta}, \\
u(\Lambda ) &&\sim\Lambda^{\epsilon-2\eta}, \\
\gamma(\Lambda ) &&\sim\Lambda^{(\epsilon +\alpha /\nu )/2-\eta}, \label{eq: glambda}\\
C(\Lambda ) &&\sim\Lambda^{-\alpha/\nu},  \label{eq: clambda}
\end{eqnarray}
where $\epsilon=4-d$ with $d$ being the space dimension, while $\alpha$, $\nu$ and $\eta$ 
are the usual static critical exponents.
Noting that $\eta$ is of order $\epsilon^2$ and very small, 
 we neglect $\eta$ and set $K_0=1$, hereafter.

\subsection{Streaming terms and  bare kinetic coefficients}

In this subsection, we  determine the forms of the streaming terms, $v_n$, $v_e$ and $\bfv_{J}$.
We can nicely determine the first two terms from the continuity equations 
because $\delta n$ and $\delta e$ are the conserved densities.
From the continuity equations, we can write $v_n$ and $v_e$ as divergences of reversible currents,
which read
\begin{eqnarray}
\bfj_{n}=&&n\gamma \delta \bfv , \label{eq: jrn}\\
\bfj_{e}=&&(e+P)\gamma^2\delta\bfv ,\label{eq: jre}
\end{eqnarray}
with $\bfj_{n}$ and $\bfj_{e}$ being the reversible currents of the number  and energy density, respectively. 
Here, $\gamma$ is the Lorentz factor of the fluid-velocity fluctuation, $n=n_c+\delta n$ and $e=e_c+\delta e$.
As the reference frame, we have chosen the rest frame of the equilibrium state,
 and then the back ground fluid velocity vanishes.
Furthermore, We may set $\gamma \sim 1$,
because the fluid-velocity fluctuation is given by $\delta \bfv = h_c^{-1} \delta \bfJ$ that is not enhanced.
Therefore, we  write the streaming terms, $v_n$ and $v_e$, as
\begin{eqnarray}
v_n&&=-\nabla \cdot (n\delta \bfv), \label{eq: vn}\\
v_e&&=-\nabla \cdot ((e+P_c)\delta \bfv ) \label{eq: ve},
\end{eqnarray}
where we neglect the pressure fluctuation because
it is not enhanced near the CP \cite{Minami:2009hn}.

Now, we note that the determination of $\bfv_{J}$ is not simple.
 Although the continuity equation tells us that 
 $\bfv_J$ is the divergence of the reversible-stress tensor,
 the determination of the reversible-stress tensor is not trivial. 
However,
we can determine it from the potential condition, 
which is a general condition for the streaming terms \cite{onuki}.  
The potential (or divergence) condition \cite{onuki, mazenko} reads
\begin{equation}
\int d\bfr \sum\limits_{j=n, e, J}  v_j(A) \frac{\delta(\beta H)}{\delta A_j }
 =\int d\bfr \sum\limits_{j=n, e, J}\frac{\partial v_j(A )}{\partial A_j }. \label{eq: potentialcon}
\end{equation}
We remark that this condition can be derived
 from the definition of streaming terms \cite{mori:fujisaka}:
\begin{equation}
 v_j(A(t)) \equiv \la \dot{A}_j ; A(t)\ra,
\end{equation}
where $\dot{A}_j \equiv i{\cal L} A_j(0)$ is the microscopic time derivative of $A_j$ and $i {\cal L}$ is the Liouville operator.
In a continuum system, the right-hand side of Eq. (\ref{eq: potentialcon})  vanishes in general \cite{onuki}. 
Thus,  the potential condition is reduced to 
\begin{equation}
\int d\bfr \sum\limits_{j=n, e, J}  v_j(A) \frac{\delta(\beta H)}{\delta A_j}=0, \label{eq: condition}
\end{equation}
where  $\bfv_J$ is only the unknown quantity 
because we have already determined $v_n$, $v_e$ and $H(\delta n, \delta e, \delta \bfJ)$.
Using Eqs. (\ref{eq: vf}), (\ref{eq: vn}), (\ref{eq: ve}) and (\ref{eq: condition}),
we obtain
\begin{eqnarray}
\int d\bfr [n\nabla \frac{\delta H}{\delta n}+(e+P_c)\nabla\frac{\delta H}{\delta e}+\bfv_J]\cdot \beta \delta\bfv=0.
\end{eqnarray}
Since this condition should be satisfied for an arbitrary fluid-velocity fluctuation, 
we have 
\begin{equation}
\bfv_J =-n\nabla\frac{\delta H}{\delta n} - (e + P_c)\nabla \frac{\delta H}{\delta e}.
\end{equation}

Next, let us determine the kinetic coefficients from 
the relativistic hydrodynamic equation, Eqs. 
(\ref{eq: n})-(\ref{eq: d2}).
From Eqs. (\ref{eq: d1}), (\ref{eq: d2}), (\ref{eq: muf}) and (\ref{eq: vf}), 
we can read the kinetic coefficients $L_{j k} $ for small $\delta \bfv$ as
\begin{eqnarray}
L_{n n}&&=-\lambda_0\bl(\frac{n_c T_c}{h_c}\br)^2\nabla^2, \\
L_{J J}^{i j}&&=-T_c [\eta_0 \delta_{i j}\partial_i \partial_j
              +(\zeta_0 +(1-2/d)\eta_0)\partial_i\partial_j ],
\end{eqnarray}
and that the other coefficients are zero.
We note that $L_{e e}$ is absent due to the choice of the energy frame. 

Now,  we have determined all the necessary terms, 
and then can write down the nonlinear Langevin equation for the QCD CP as 
\begin{eqnarray}
\frac{\partial \delta n}{\partial t}=&&-\nabla \cdot (n\delta \bfv)
                  -L_{n n}\frac{\delta (\beta H)}{\delta n}+\theta_n , \label{eq: nln} \\
\frac{\partial \delta e}{\partial t}=&&-\nabla \cdot ((e+P_c)\delta \bfv) ,\label{eq: nle} \\
\frac{\partial \delta \bfJ}{\partial t}=&&-n\nabla \frac{\delta H}{\delta n}
-(e+P_c)\nabla\frac{\delta H}{\delta e} \nonumber \\
&&-L_{J J} \cdot \frac{\delta(\beta H)}{\delta \bfJ}+ \bftheta_J ,\label{eq: nlj}
\end{eqnarray}
where $\theta_n$ and $\bftheta_J$ are the noise terms and satisfy the fluctuation-dissipation relations
\begin{eqnarray}
\la \theta_n (\bfr ,t) \theta_n (\bfr^{'} ,t^{'})\ra =&&-2\lambda_{0}
    \bl(\frac{n_c T_c}{h_c}\br)^2\nabla^2\delta (\bfr -\bfr^{'} ) \nonumber \\
  &&\times \delta (t-t^{'}), \label{eq: fdrn} \\
\la \theta_J^i(\bfr,t) \theta_J^i(\bfr^{'},t^{'}) \ra =&&-2 T_c[\eta_0 \delta^{i j}\nabla^2 \nonumber \\
   &&+\{\zeta_0+(1-2/d)\eta_0\}\partial^i\partial^j] \nonumber \\
   &&\times\delta(\bfr-\bfr^{'})\delta(t-t^{'}). \label{eq: fdrj}
\end{eqnarray}

Let us write the transport coefficients as $\lambda(\Lambda)$, $\eta(\Lambda)$ and $\zeta(\Lambda)$
to make their cutoff dependence in the region, $\xi^{-1} < \Lambda$.
The critical behaviors of the transport coefficients
 are determined from their asymptotic behaviors near the relevant fixed point
as  $\Lambda$ is lowered.  

Here, we compare the Langevin equation Eqs. (\ref{eq: nln}) - (\ref{eq: nlj}) 
with that for the liquid-gas CP \cite{onukin}
\begin{eqnarray}
\frac{\partial \delta n}{\partial t}=&&-\nabla \cdot (n\delta v),\\
\frac{\partial \delta e}{\partial t}=&&-\nabla \cdot ((e+P_c)\delta v)
  +\lambda_0 T_c \nabla^2\frac{\delta H^{\rm lg}}{\delta e}+\theta_e ,\\
\frac{\partial \delta \bfJ_{\rho}}{\partial t}=&&-n\nabla \frac{\delta H^{\rm lg}}{\delta n}
  -(e+P_c)\nabla\frac{\delta H^{\rm lg}}{\delta e} \nonumber \\
  &&-L_{J J} \cdot \frac{\delta(\beta H^{\rm lg})}{\delta \bfJ_\rho}+ \bftheta_J ,
\end{eqnarray}
where $\delta \bfJ_\rho \equiv \rho_c \delta \bfv$, 
$\rho$ and $H^{\rm lg}$ are  the non-relativistic momentum density, the mass density
 and the thermodynamic potential for liquid-gas CP, respectively:  
\begin{equation}
H^{\rm lg}(\delta n,\delta e, \delta \bfJ )=H_{\rm Ising}(\psi, m) + \frac{1}{2 \rho_c}\delta \bfJ^2.  
\end{equation}
We see that the streaming terms have the same forms but
the dissipative ones are totally different between the relativistic and non-relativistic cases.
The difference also appears the relation between the momentum
and the fluid-velocity fluctuation.
Therefore, one may naturally  expect some novel characteristics  in the 
relativistic case that is absent in the non-relativistic case \cite{onukin}.

\section{The transport coefficients by dynamic RG }

We here present an analysis of transport coefficients by the dynamic RG.
A detailed derivation of the RG equations is given in the Appendixes.

First, we rewrite Eqs. (\ref{eq: nln}) - (\ref{eq: nlj}) as the equation for $\psi$ and $m$
to conform the hydrodynamic variables, $\delta n$ and $\delta e$, to the Ising variables, $\psi$ and $m$.
Noting that we can set $\alpha_2=0$ in the mapping relations,
 Eqs. (\ref{eq: map1}) and (\ref{eq: map2}), without loss of generality \cite{onukin},
we have
\begin{eqnarray}
\frac{\partial \psi}{\partial t}
    =&&-C_{\psi} \nabla\cdot\delta \bfJ -\alpha_1^{-2}L_{n n}\frac{\delta(\beta H)}{\delta \psi} \nonumber \\
      &&-h_c^{-1}\nabla\cdot (\psi \delta \bfJ) +\alpha_1^{-1}\theta_n ,\label{eq:psi} \\
\frac{\partial m}{\partial t} =&&-\beta_2^{-1}\nabla\cdot\delta \bfJ -h_c^{-1}\nabla\cdot(m\delta \bfJ) ,
  \label{eq:m} \\
\frac{\partial(\delta \bfJ)}{\partial t}
    =&&
      -C_J\nabla\frac{\delta H}{\delta \psi}-\beta_2^{-1}h_c\nabla\frac{\delta H}{\delta m}
     \nonumber \\
      &&-\psi \nabla \frac{\delta H}{\delta \psi}-m\nabla\frac{\delta H}{\delta m} 
     -(T_c h_c)^{-1}L_{J J}\cdot \delta \bfJ \nonumber \\
     && +\bftheta_J ,
\label{eq:J}
\end{eqnarray}
with
$C_{\psi} \equiv \alpha_1^{-1}(n_c h_c^{-1}-\beta_1\beta_2^{-1})$ and
$C_{J}\equiv \alpha_1^{-1}(n_c-\beta_1 h_c)$. 
Here, we note that we could rewrite the potential, Eq. (\ref{eq: glw}), for $\psi$ and $m$ 
as that for $\delta n$ and $\delta e$ to conform the variables;
the choice is a matter of preference.

In the dynamic RG transformation, 
we average over the short wavelength components in the shell, 
$\Lambda-\delta\Lambda < k < \Lambda$, for the Langevin equation.
For this task, we must perturbatively solve the equation about them,
by rewriting it as a self-consistent equation \cite{mazenko}.    
Although an explicit derivation of the self-consistent equation for the QCD CP is first made in this paper,
 we leave the details of the derivation to Appendix \ref{sec: derivation},
because the general procedure of the derivation is
standard and given in the textbook \cite{mazenko}.
Here, we present only a few basic equations of the dynamic RG for the QCD CP.
Now, as is shown in Appendix \ref{sec: derivation},
 Eqs. (\ref{eq:psi})-(\ref{eq:J}) can be reduced to the following form; 
\begin{equation}
  \begin{pmatrix}
  \tilde{\psi}(\bfk ,\omega) \\
  \tilde{m}(\bfk ,\omega) \\
  \delta \tilde{J}_{\parallel}(\bfk,\omega)
  \end{pmatrix}
=
 \begin{pmatrix}
  \tilde{\psi}^0(\bfk ,\omega) \\
  0 \\
  \delta \tilde{J}_{\parallel}^0(\bfk,\omega)
  \end{pmatrix}
+G^0(\bfk,\omega)\bfV(\bfk,\omega),
\label{eq: self1}
\end{equation}
and
\begin{equation}
\delta \tilde{\bfJ}_{\perp}(\bfk,\omega)
=\delta \tilde{\bfJ}_{\perp}^0(\bfk,\omega)+G^0_{\perp}\bfV_{\perp \psi\psi}(\bfk,\omega),
\label{eq: self2}
\end{equation}
where $\tilde{J}_{\parallel }(\bfk) \equiv \hat{\bfk} \cdot \tilde{J}(\bfk)$ and 
$\tilde{J}_{\perp}(\bfk) \equiv \tilde{J}(\bfk) - \tilde{J}_{\parallel}(\bfk)$ are 
the longitudinal and  transverse components of the momentum.  
Here, $G^0$ and $G^0_{\perp}$ are the bare propagators, which are given by Eqs. (\ref{eq: g0}) and 
 (\ref{eq: gthermal}) - (\ref{eq: gviscous}),
whereas $\bfV$ and $\bfV_{\perp \psi\psi}$ the nonlinear couplings, coming from the streaming terms
 and given by Eqs. (\ref{eq: v1}) - (\ref{eq: v3}) and (\ref{eq: v4}). 
Also, $\tilde{\psi}^0$, $\delta \tilde{J}_{\parallel}^0$ and $\delta \tilde{\bfJ}_{\perp}^0$
are the bare variables, which are the solutions without the nonlinear terms.  
Iterating the self-consistent equations (\ref{eq: self1}) and (\ref{eq: self2}), 
we can obtain a perturbative expansion of the nonlinear couplings
and have a coarse-grained Langevin equation.

Now, we note that the variables, $\psi $,$\tilde{J}_{\perp}$ and $\tilde{J}_{\parallel} $,
are respectively correspond to the thermal,  viscous and  sound modes
\footnote{Although $\tilde{m}$ would be a linear combination of the thermal and  sound modes,
we need not to consider $\tilde{m}$ for a following analysis. } 
(see the propagators  (\ref{eq: gthermal}) - (\ref{eq: gviscous}).).
Therefore, the first and third rows of Eq. (\ref{eq: self1}) 
respectively denote the equations of  motion for the thermal and sound modes,
while Eq. (\ref{eq: self2})  for the viscous mode. 
We stress that the sound mode is neglected in the model H,
although it is essential for the renormalization of the bulk viscosity.    

Here, we  make the coarse graining 
to the second order in the nonlinear couplings,  $\bfV$ and $\bfV_{\perp \psi\psi}$
(see Fig. \ref{fig: cgpeom} for an example.).
Inspecting the coarse-grained equation for $\tilde{\psi}$ 
(see Eq. (\ref{eq: cgpeom}) for the detail) and,
we have the RG equation for the thermal conductivity:
\begin{eqnarray}    
-\Lambda\frac{\partial \lambda(\Lambda)}{\partial \Lambda}=&&\frac{3}{4}f(\Lambda)\lambda(\Lambda),
\label{eq: rgl} 
\end{eqnarray}
$f(\Lambda)\equiv T_c K_4/(D_{\psi}\eta(\Lambda)\lambda(\Lambda)\Lambda^{\epsilon}) $,
$K_4$ is the surface area of a unit sphere in 4 dimensions divided by $(2\pi)^4$,
$D_{\psi}\equiv (n_c T_c /\alpha_1 h_c)^2$.
Here, we have introduced $f(\Lambda)$ for convenience sake.
Similarly, from the coarse-grained equations for $\tilde{J}_{\perp} $ and $\tilde{J}_{\parallel} $,
we obtain the RG equations for the shear and bulk viscosities
\begin{eqnarray}  
-\Lambda\frac{\partial \eta(\Lambda)}{\partial \Lambda}=&&\frac{1}{24}f(\Lambda)\eta(\Lambda),
\label{eq: rge} \\
-\Lambda\frac{\partial \zeta(\Lambda)}{\partial \Lambda}=&&
A\gamma^2(\Lambda)\lambda^{-1}(\Lambda)\Lambda^{-\epsilon-4},
\label{eq: rgz}
\end{eqnarray}
where  $\gamma(\Lambda)$ is a static parameter in the thermodynamic potential 
(see Eqs. (\ref{eq: glw}) and (\ref{eq: glambda})), and $A \equiv  h_c^2 K_4/(\beta_2^2 D_{\psi})$.
Furthermore, differentiating $f(\Lambda)$ about $\Lambda$,
we also have the RG equation for it:
\begin{equation}
-\Lambda\frac{\partial f(\Lambda)}{\partial \Lambda}=f(\Lambda )(\epsilon-\frac{19}{24}f(\Lambda )).
\label{eq: rgf} 
\end{equation}

Now, we note that Eqs. (\ref{eq: rgl}), (\ref{eq: rge}) and (\ref{eq: rgf}) are identical to those for  the
liquid-gas CP except for unimportant constants in $f(\Lambda)$\cite{onuki, onukin}.
Equation (\ref{eq: rgz}) is also equivalent 
to the RG equation of the bulk viscosity for the liquid-gas CP in the limit $\omega \rightarrow 0$ 
\cite{onuki, onukin}. 
Therefore, arguments about the RG equations and results from those are the same as for the liquid-gas CP.
Then, we provide only essential  arguments and results in the following part,
and leave the detail to \cite{onuki, kg, siggia, onukin}.

Now, we identify the relevant-fixed point as the following \cite{onuki, onukin}.
Because, at a fixed point, parameters are invariant about the RG transformation,
we set the left-hand side of Eq. (\ref{eq: rgf}) as $0$.
Then, as a fixed-point value of $f(\Lambda)$ which is denoted by $f^{*}$,
 we have $f^*=0$ and $f^{*}=(24/19)\epsilon$.
Therefore, we have the two fixed points and the relevant one is specified by $f^{*}=(24/19)\epsilon$.
Although the relevant point seems to be absent in Eqs. (\ref{eq: rgl}), (\ref{eq: rge}) and (\ref{eq: rgz}),
 the reason is due to the simplified RG transformation as mentioned in the earlier section,
and this is just a apparent problem \cite{kg, siggia}.   

Substituting $f^{*}=(24/19)\epsilon$ into Eqs. (\ref{eq: rgl}), (\ref{eq: rge}) and (\ref{eq: rgz}),
we have the asymptotic behaviors near the relevant-fixed point:
\begin{eqnarray}
\lambda(\Lambda) \sim && \Lambda^{-\frac{18}{19}\epsilon}, \\
\eta(\Lambda) \sim && \Lambda^{-\frac{1}{19}\epsilon}, \\
\zeta(\Lambda) \sim && \Lambda^{-(4-\frac{18}{19}\epsilon-\frac{\alpha}{\nu})} \label{eq: zlambda}.
\end{eqnarray} 
Here, in the derivation of Eq. (\ref{eq: zlambda}), we have used the asymptotic behavior of $\gamma(\Lambda)$,
 Eq. (\ref{eq: glambda}).
Decreasing the cutoff to the region $\Lambda \ll \xi^{-1}$, 
we can replace $\Lambda$ with $ \xi^{-1}$ in the asymptotic behaviors \cite{onuki, siggia}:
\begin{eqnarray}
\lambda_{\rm R}\sim && \xi^{\frac{18}{19}\epsilon}, \label{eq: lr}\\
\eta_{\rm R} \sim && \xi^{\frac{1}{19}\epsilon}, \label{eq: er}\\
\zeta_{\rm R} \sim && \xi^{4-\frac{18}{19}\epsilon-\frac{\alpha}{\nu}}. \label{eq: zr}
\end{eqnarray} 
In three dimensions, we find
\begin{eqnarray}
\lambda_{\rm R}\sim && \xi^{0.95}, \\
\eta_{\rm R} \sim && \xi^{0.053}, \\
\zeta_{\rm R} \sim && \xi^{2.8}.
\end{eqnarray} 

We can also read the dynamic critical exponents from Eqs. (\ref{eq: lr})-(\ref{eq: zr}).
A dynamic critical exponent, denoted by $z$, generically parametrizes 
the decay rate $\Gamma(k)$ at the wavenumber $k=\xi^{-1}$ as 
$\Gamma (\xi^{-1}) \sim \xi^{-z}$.
As shown in Appendix \ref{sec: derivation},
the decay rates for the three modes 
at $k$ are given by
\begin{eqnarray}
\Gamma_{\rm thermal}(k) =&& \lambda_{\rm R} k^2 (r_{\rm R} +k^2) D_{\psi}, \\
\Gamma_{\rm viscous}(k) =&& \eta_{\rm R} k^2 h_c^{-1}, \\
\Gamma_{\rm sound}(k) =&&  (\zeta_{\rm R} +2(1-1/d)\eta_{\rm R}) k^2 h_c^{-1}.
\end{eqnarray}  
Thus, we find the dynamic critical exponents as
\begin{eqnarray}
z_{\rm thermal} =&&4-\frac{18}{19}\epsilon, \\
z_{\rm viscous} =&&2-\frac{1}{19}\epsilon, \\
z_{\rm sound} =&&-(2-\frac{18}{19}\epsilon-\frac{\alpha}{\nu} ).
\end{eqnarray}
In three dimensions, the dynamic critical exponents are given by
\begin{eqnarray}
z_{\rm thermal} \sim &&3, \\
z_{\rm viscous} \sim &&2, \\
z_{\rm sound} \sim &&-0.8.
\end{eqnarray}
We see that the thermal and  viscous modes exhibit critical slowing down, while
the sound mode critical speeding up.

Why do not the relativistic effects appear in the RG equations?
The reason is that
the nonlinear terms in the dissipative terms 
generally renormalize only static parameters, 
up to order $\epsilon^2$ \cite{siggia, mazenko}.
Furthermore,
the difference in the relation between the momentum
and the fluid velocity is only unimportant constants, 
i.e., the enthalpy density $h$ and the mass density $\rho$.
Then, the RG equations are essentially the same as for the non-relativistic case.

\section{Summary and Concluding remarks}

We have studied the critical behaviors of the transport coefficients 
and the dynamic critical exponents at the QCD critical point (CP) by dynamic renormalization group (RG).
For this purpose, we have constructed
 the nonlinear Langevin equation  near  the QCD CP
for the first time. 
Our construction is 
based on the generalized Langevin theory, by 
Mori \cite{mori,mori:fujisaka}, and the relativistic hydrodynamics;
instead of a naive construction method \cite{mazenko},
we have determined  the
streaming terms by  the relativistic hydrodynamics
and the potential condition that gives a constraint to these terms.  
The resulting equation is given by Eqs. (\ref{eq: nln})-(\ref{eq: nlj}). 
Although there are some attempts to 
make a one-to-one mapping between QCD CP and Ising CP \cite{moore, mapping},
we have shown that it  is not necessary to specify such the mapping for the 
critical exponents, as for the liquid-gas CP \cite{onuki}.    

We have shown that 
the bulk viscosity and the thermal conductivity strongly diverge  at the QCD CP. 
Also, we have found that the thermal and viscous diffusion modes 
exhibit critical slowing down with the dynamic critical
exponents $z_{\rm thermal}\sim 3$  and $z_{\rm viscous}\sim 2$, respectively.  
In contrast,  the sound propagating mode critical speeding up
with the negative exponent $z_{\rm sound}\sim -0.8$.

We now compare our result about the bulk viscosity to that in \cite{karsch}.
Although a divergent behavior of the bulk viscosity is the same,
the critical exponents is different in the two cases.
In \cite{karsch}, the critical exponent is estimated to be about $0.2$
and the divergence is weak contrary to our result.
We also note that the study by the relativistic Boltzmann equation \cite{sasaki}
 gives only the bare bulk viscosity.

We note that the bulk viscosity and the thermal conductivity
are usually neglected in  heavy ion physics, however
they become much more important than the shear viscosity near the QCD CP.
Furthermore,  the description for the created matter as a perfect fluid is
not valid near the QCD CP at all
due to the strong divergence of the bulk viscosity.

As the argument about the dynamic universality class \cite{moore, son},
we have shown, from an explicit calculation, 
that the QCD CP has the same critical behaviors as the liquid-gas CP has. 
The argument assumes the insignificance of the relativity for the critical dynamics by
 the slowness of the diffusion processes.
However, 
we have shown that the genuine reason for the insignificance
originates from the small fluctuation of the momentum density;
the critical dynamics is essentially governed by the streaming terms,
which are modified by  the relativistic effect
through only a Lorentz factor of the fluid velocity fluctuation.
However, the fluid velocity fluctuation, which is proportional to the momentum, 
is not enhanced near the CP.
Thus, the relativistic effects do not affect the critical 
dynamics near the QCD CP.
We stress that the sound mode exhibit critical speeding up,
and then the sound diffusion is fast near the QCD CP. 
Therefore, the basis of the conjecture would be true for the thermal and  viscous modes,
but not for the sound mode.
We also note that the model H \cite{kawasaki1},
which is the {\em minimal}-dynamic model for the dynamics near the liquid-gas CP,
 can not describe the critical behavior of the bulk viscosity 
because it does not contain the sound mode.

We note that our Langevin equation must satisfy usual fluctuation-dissipation relations, 
Eqs. (\ref{eq: fdrn}) and (\ref{eq: fdrj}), for the consistency with the linearized Landau equation
\footnote{If our nonlinear Langevin equation is linearized, 
the linearized equation must give the same result as the Landau equation gives.},
although a relativistic Brownian motion seems not  to satisfy the usual relations \cite{Akamatsu:2008ge}.
Moreover, our Langevin equation seems to violate the causality, 
since the dissipative terms are determined from the Landau equation.
However,  the Israel-Stewart equation \cite{is}, in which the causality problem is formally resolved,
 gives the same result as the Landau equation gives in long-wavelength region \cite{Minami:2009hn}. 
Therefore, our determination from the Landau equation must suffice.
Furthermore, we note that short-wavelength components in the region, 
$k > a^{-1}$ where $a$ is a characteristic microscopic length, would violate the causality.
Therefore, we can exclude such the illegal components from the theory by the cutoff, $\Lambda_0 \ll a^{-1}$.    
We stress that all infrared effective theories inevitably have the ultraviolet cutoff; 
naturally, relativistic hydrodynamics also has it.

Also, we note a frame dependence of the results.
As a  hydrodynamic equation, we used only the equation in the energy frame.
Does the results change if an equation in the particle frame is used?
Although the frame dependence can appear in only dissipative terms,
 the critical dynamics is essentially determined by the streaming terms.
Therefore, the results would not change for the particle frame, if an equation in the frame is correct. 
However, in practice, the Eckart equation has a pathological behavior\cite{hiscock}.
Namely, fluctuations do not relax, and therefore we cannot use the Eckart equation.

Recently, some authors have suggested the existence of other critical points in higher density region
of the QCD phase diagram where the color superconductivity is taken into account
\cite{Kitazawa:2002bc,Hatsuda:2006ps}.
It would be interesting to study the critical dynamics near  such a new QCD CP using the dynamic RG theory,
as an extension of the present work.
For this purpose, however, we must firstly specify the soft modes and 
construct the nonlinear Langevin equation.
If the soft modes are different from the conserved densities, which is the case when
the the diquark fluctuations are relevant \cite{Kitazawa:2002bc,Kitazawa:2001ft},
the construction based on the relativistic hydrodynamics done in the present work does not work,
and we must directly recourse to  Eq. (\ref{eq: stream}) to identify  the streaming terms.    

\section*{ACKNOWLEDGMENTS}
We are grateful to Hideo Suganuma for his useful comments.
We also thank Teiji Kunihiro for his careful reading this paper.
This work was supported by 
 the Global COE Program ``The Next Generation of Physics, Spun
from Universality and Emergence'' in Kyoto University, by the Yukawa
International Program for Quark-hadron Sciences in YITP and by the
Grant-in-Aid for Scientific Research in Japan [Nos. 22-1050].

\appendix

\section{Rewriting the nonlinear Langevin equation as a self-consistent equation}
\label{sec: derivation}
Here, we rewrite the Langevin equation, Eqs. (\ref{eq:psi})-(\ref{eq:J}) as a self-consistent equation.
First, we make a Fourier transformation as the following
\begin{eqnarray}
\tilde{\psi}(\bfk,\omega) =\int dt d^d r e^{i\omega t-i\bfk\cdot\bfr} \psi(\bfr,t). \label{eq: foo}
\end{eqnarray}
Then, we have
\begin{eqnarray}
-i\omega\tilde{\psi}(\bfk,\omega)
 =&&- C_{\psi} i\bfk \cdot\delta\tilde{\bfJ}
    -\alpha_1^{-2}\tilde{L_{n n}}\frac{\delta(\beta \tilde{H})}{\delta \psi} \nonumber \\
    &&-h_c^{-1} i\bfk \cdot \int_{q \Omega} (\tilde \psi(q) \delta \tilde{\bfJ}(k-q)) +\alpha_1^{-1}\tilde{\theta_n}, 
  \label{eq: fpsi}\\
-i\omega \tilde{m}(\bfk,\omega)
  =&&-\beta_2^{-1}i\bfk\cdot\delta \tilde{\bfJ} \nonumber \\ 
    &&-h_c^{-1}i\bfk\cdot\int_{q\Omega}(\tilde{m}(q)\delta \tilde{\bfJ}(k-q)), \\
-i\omega\delta\tilde{\bfJ}(\bfk,\omega ) 
 =&&
 -C_J i\bfk\frac{\delta \tilde{H}}{\delta \psi}-\beta_2^{-1}h_c i\bfk\frac{\delta \tilde{H}}{\delta m}
 \nonumber \\
    &&-i\int_{q\Omega}\bfq  [\frac{\delta \tilde{H}}{\delta \psi}(q)\tilde{\psi}(k-q)
                                +\frac{\delta \tilde{H}}{\delta m}(q)\tilde{m}(k-q)] \nonumber \\ 
    &&-(T_c h_c)^{-1}\tilde{L}_{J J}\cdot \delta \tilde{\bfJ} +\tilde{\bftheta_J}. \label{eq: fj}
\end{eqnarray}
Note that the quantities with tilde in Eq. (\ref{eq: fpsi})-(\ref{eq: fj}) are Fourier transformed,
like Eq. (\ref{eq: foo}), and we have abbreviated the nonlinear terms such as 
\begin{equation}
\int_{q\Omega}\tilde{\psi}(q)\delta\tilde{\bfJ}(k-q)
\equiv \int\frac{d \Omega}{2\pi}\frac{d^d q}{(2\pi)^d}\tilde{\psi}(\bfq,\Omega)\delta\tilde{\bfJ}(\bfk-\bfq,\omega-\Omega).
\end{equation}
We now decompose Eq. (\ref{eq: fj}) into the longitudinal and the transverse components:
\begin{eqnarray}
-i\omega\delta\tilde{\bfJ}_{\parallel}
   =&&-i C_J k \frac{\delta \tilde{H}}{\delta \psi}-i\beta_2^{-1}h_c k\frac{\delta \tilde{H}}{\delta m} \nonumber \\
   &&-i\int_{q\Omega}(\hat{\bfk}\cdot\bfq)  \nonumber \\
              &&                \times [\frac{\delta \tilde{H}}{\delta \psi}(q)\tilde{\psi}(k-q)
                                +\frac{\delta \tilde{H}}{\delta m}(q)\tilde{m}(k-q)]  \nonumber \\
   &&-(T_c H_c)^{-1}\hat{\bfk}\cdot\tilde{L}_{J J}(\bfk )\cdot\delta\tilde{\bfJ}+\tilde\theta_{\parallel} \label{eq: parallel}, \\
-i\omega\delta \tilde{\bfJ}_{\perp}
   &&=-i\int_{q\Omega}{\cal P}_{\perp}(\bfk) \cdot\bfq  \nonumber \\
              &&                \times [\frac{\delta \tilde{H}}{\delta \psi}(q)\tilde{\psi}(k-q)
                                +\frac{\delta \tilde{H}}{\delta m}(q)\tilde{m}(k-q)]\nonumber \\ 
   &&   -(T_c h_c)^{-1}{\cal P}_{\perp}(\bfk) \cdot \tilde{L}_{J J}(\bfk) \cdot \delta \tilde{\bfJ} \nonumber \\
   &&+\tilde{\bftheta}_{\perp} \label{eq: perp},
\end{eqnarray}
where we have introduced a projection operator  as
\begin{equation}
({\cal P}_{\perp}(\bfk))_{i j}=\delta_{i j }-k_i k_j /k^2,
\end{equation}
and
\begin{eqnarray}
\delta \tilde{J}_{\parallel}(\bfk) &&=\hat{\bfk} \cdot \delta \tilde{\bfJ}(\bfk), \\
\delta \tilde{\bfJ}_{\perp}(\bfk) &&={\cal P}_{\perp}(\bfk) \cdot \delta \tilde{\bfJ} (\bfk), \\
\tilde{\theta}_{\parallel}(\bfk) &&=\hat{\bfk}\cdot\tilde{\bftheta}(\bfk),\\
\tilde{\bftheta}_{\perp}(\bfk) &&= {\cal P}_{\perp}(\bfk) \cdot \tilde{\bftheta}(\bfk).
\end{eqnarray}
Because the streaming terms in Eqs. (\ref{eq: parallel}) and (\ref{eq: perp}) are too complicated
for our purpose,
let us retain only the terms that yield dominant contributions for the transport coefficients.
We note that only such terms suffice for obtaining the critical exponents.
From the relations\cite{onuki} 
\begin{eqnarray}
\int d^3 r \la\psi(\bfr)\psi(0) \ra &&\sim \xi^2 ,\\
\int d^3 r \la m(\bfr) m(0) \ra && \sim \xi^{0.2} ,
\end{eqnarray}
we expect $\psi $ yields stronger singularity than $m$.
Therefore, we only retain the term that are of the second order in $\psi$.
Namely, we reduce the streaming terms as
\begin{eqnarray}
&&i C_J k \frac{\delta \tilde{H}}{\delta \psi}+i\beta_2^{-1}h_c k\frac{\delta \tilde{H}}{\delta m}
\nonumber \\
     && \hspace{0.5cm} +i\int_{q\Omega}(\hat{    \bfk}\cdot\bfq)  [\frac{\delta \tilde{H}}{\delta \psi}(q)\tilde{\psi}(k-q)
    +\frac{\delta \tilde{H}}{\delta m}(q)\tilde{m}(k-q)]
     \nonumber \\
     &&\sim T_c[ i C_J k\chi_0^{-1}(k)\tilde{\psi}+i\beta_2^{-1}h_c k C_0^{-1}\tilde{m}
     \nonumber \\
     &&\hspace{0.5cm} +i\beta_2^{-1}h_c k\gamma_0 \int_{q\Omega}\tilde{\psi}(q)\tilde{\psi}(k-q)], \\
&&i\int_{q\Omega}{\cal P}_{\perp}(\bfk) \cdot\bfq  [\frac{\delta \tilde{H}}{\delta \psi}(q)\tilde{\psi}(k-q)
     +\frac{\delta \tilde{H}}{\delta m}(q)\tilde{m}(k-q)] \nonumber \\
     &&\sim  i T_c{\cal P}_{\perp}(\bfk)\cdot\int_{q\Omega} \bfq \chi_{0}^{-1}(\bfq) \tilde{\psi}(q)\tilde{\psi}(k-q), 
\end{eqnarray}
where $\chi_0^{-1}(\bfk)=r_0+ k^2 $.
Notice that we have set  $K_0=1$, as mentioned in the text.

Next, we consider the dissipative terms.
The important point is that the nonlinear terms in dissipative terms renormalize 
 only static parameters in a thermodynamic potential to second order in $\epsilon$,
 generally \cite{siggia, mazenko}.
Therefore,
we can take into account  nonlinear terms in the dissipative terms
with the results of static RG, Eq. (\ref{eq: rlambda})-(\ref{eq: clambda}),
and effectively neglect it in the Langevin equation.
Then, we reduce the $\tilde{L}_{n n}\delta(\beta \tilde{H})/\delta \psi$ as
\begin{equation}
\tilde{L}_{n n}(\bfk )\frac{\delta \tilde{H}}{\delta \psi}(\bfk ,\omega)
\sim \lambda_0 k^2 \chi_0^{-1} \bl(\frac{n_c T_c}{h_c}\br)^2 \tilde{\psi}(\bfk,\omega).
\end{equation}
In contrast, the dissipative terms of $\delta \bfJ$ are originally linear and then directly read
\begin{eqnarray}
\hat{\bfk} \cdot \tilde{L}_{J J}(\bfk) \cdot \delta \bfJ(\bfk,\omega) 
  &&= T_c[\zeta_0+2(1-1/d)\eta_0] \nonumber \\
  &&\times k^2\delta\tilde{J}_{\parallel}(\bfk,\omega) ,\\
{\cal P}_{\perp} (\bfk)  \cdot \tilde{L}_{J J}(\bfk) \cdot \delta \bfJ(\bfk,\omega)&&
   = T_c\eta_0 k^2 \delta \tilde{\bfJ}_{\perp}(\bfk,\omega).
\end{eqnarray}

Collecting the above results, we arrive at the reduced nonlinear Langevin equation as given by
\begin{eqnarray}
-i\omega\tilde{\psi}
   =&&-i k C_{\psi}\delta\tilde{J}_{\parallel} \nonumber \\
     &&  -h_c^{-1} i\bfk \cdot \int_{q\Omega} \tilde{\psi}(q)\delta\tilde{\bfJ}(k-q) \nonumber \\
     &&-\lambda_0 k^2 D_{\psi}\chi^{-1}_0(\bfk)\tilde{\psi}+\alpha_1^{-1}\tilde{\theta}_n, \\
-i\omega\tilde{m}
   =&&-\beta_2^{-1} i k \delta \tilde{J}_{\parallel} \nonumber \\
     &&-h_c^{-1}i\bfk\cdot\int_{q\Omega}\tilde{m}(q)\delta\tilde{\bfJ}(k-q), \\
-i\omega \delta\tilde{J}_{\parallel}
   =&&T_c[-ik\chi_0^{-1}(\bfk)C_J \tilde{\psi}-ikC_0^{-1}\beta_2h_c\tilde{m} \nonumber \\
     &&-i k\beta_2^{-1}h_c\gamma_0 \int_{q\Omega}\tilde{\psi}(q)\tilde{\psi}(k-q) ] \nonumber \\
     &&-k^2\nu_0^l h_c^{-1} \delta\tilde{J}_{\parallel}+\tilde{\theta}_{\parallel}, \label{eq: soundeom} \\
-i\omega\delta\tilde{\bfJ}_{\perp}
   =&&-i T_c {\cal P}_{\perp}(\bfk )\cdot\int_{q\Omega} \bfq 
         \chi_0^{-1}(\bfq) \tilde{\psi}(q)\tilde{\psi}(k-q) \nonumber \\ 
     &&-k^2\eta_0 h_c^{-1}\delta\tilde{\bfJ}_{\perp}+\tilde{\bftheta}_{\perp}, \label{eq: transverse}
\end{eqnarray}
where 
\begin{eqnarray}
D_{\psi} &&\equiv \bl( \frac{n_c T_c}{\alpha_1 h_c}\br)^2, \\
\nu_0^{l} &&\equiv [\zeta_0+2(1-1/d)\eta_0].
\end{eqnarray}
This is the basic equation for the dynamics near the QCD CP, which is first written down,
and a main result of this paper.

We can compactly rewrite the basic equation in a matrix form:
\begin{equation}
{\cal M}(\bfk,\omega)
  \begin{pmatrix}
  \tilde{\psi}(\bfk ,\omega) \\
  \tilde{m}(\bfk ,\omega) \\
  \delta \tilde{J}_{\parallel}(\bfk,\omega)
  \end{pmatrix}
=\bfV(\bfk,\omega)+\bftheta(\bfk, \omega),
\end{equation}
where
\begin{widetext}
 \begin{equation}
{\cal M}(\bfk, \omega)= 
  \begin{pmatrix}
   -i\omega+\lambda_0 k^2 D_{\psi}\chi^{-1}_0(\bfk) & 0 & i k C_{\psi}  \\
   0 & -i\omega & i k\beta_2^{-1}\\
    i k \chi_0^{-1}(\bfk)C_J T_c &  i k C_0^{-1}\beta_2 h_c T_c & -i\omega +k^2\nu_0^{l} h_c^{-1}
  \end{pmatrix},
\end{equation}
\end{widetext}
\begin{equation}
\bftheta(\bfk,\omega)= 
  \begin{pmatrix}
  \alpha_1^{-1} \tilde{\theta}(\bfk,\omega)\\
  0 \\
  \tilde{\theta}_{\parallel}(\bfk,\omega)
  \end{pmatrix},
\end{equation}
\begin{equation}
\bfV(\bfk, \omega)= 
  \begin{pmatrix}
  V_{\psi\psi\perp}(\bfk,\omega) +V_{\psi\psi\parallel}(\bfk,\omega) \\
  V_{m m\perp}(\bfk,\omega) +V_{m m\parallel}(\bfk,\omega) \\
  V_{\parallel\psi\psi}(\bfk,\omega)
  \end{pmatrix}, \label{eq: v1}
\end{equation}
and
\begin{eqnarray}
V_{\psi\psi\perp}(\bfk,\omega) &&\equiv
    - h_c^{-1} i\bfk \cdot \int_{q\Omega} \tilde{\psi}(q)\delta\tilde{\bfJ}_{\perp}(k-q), \label{eq: v2 }\\
V_{\psi\psi\parallel}(\bfk,\omega) && \equiv 
    -h_c^{-1} i   \int_{q\Omega} \bfk\cdot(\bfk-\bfq)/|\bfk-\bfq| 
   \nonumber \\ && \times \tilde{\psi}(q)\delta\tilde{J}_{\parallel}(k-q), \\
V_{m m\perp}(\bfk,\omega) &&\equiv -h_c^{-1}i\bfk\cdot\int_{q\Omega}\tilde{m}(q)\delta\tilde{\bfJ}_{\perp}(k-q), \\
V_{m m\parallel}(\bfk,\omega) &&\equiv -h_c^{-1}i k\int_{q\Omega}\tilde{m}(q)\delta\tilde{J}_{\parallel}(k-q), \\
V_{\parallel\psi\psi}(\bfk,\omega) &&\equiv 
\hspace{-0.1cm}-i k T_c\beta_2^{-1}h_c\gamma_0 \int_{q\Omega}\hspace{-0.1cm}
\hspace{-0.2cm}\tilde{\psi}(q)\tilde{\psi}(k-q).\label{eq: v3}
\end{eqnarray}
Since Eq. (\ref{eq: transverse}) is decoupled from the other equations at linear level, 
we do not rewrite it as the matrix form.

Next, we calculate the bare propagator $G^{0}(\bfk ,\omega)={\cal M}^{-1}(\bfk, \omega)$.
The inverse matrix is given as the transposed cofactor matrix divided by $\det {\cal M}$.
The determinant reads
\begin{eqnarray}
\det{\cal M}=&&(-i\omega)^3+(-i\omega)^2 k^2(\lambda D_{\psi}\chi_0^{-1}(\bfk )+\nu_0^l h_c^{-1} ) \nonumber \\
  &&-i\omega k^2(C_0^{-1}h_c T_c+\chi_0^{-1}(\bfk )C_{\psi}C_J T_c) \nonumber \\
  &&+k^4\lambda_0\chi_0^{-1}D_{\psi}C_0^{-1}h_c T_c \nonumber \\
  &&-i\omega k^4\lambda_0\chi_0^{-1}D_{\psi}(\bfk) \nu_0^l h_c^{-1}.
\end{eqnarray}
Here, in the coefficient of  $-i\omega k^2$, 
taking into account the behaviors after renormalization \cite{onukin, glw}, which are given as  
\begin{eqnarray}
C_R^{-1} &&\sim \xi^{-0.2}, \\
\chi_R^{-1}(\bfk ) &&\sim \xi^{-2}+k^2,
\end{eqnarray}
 we neglect $\chi_0^{-1}(\bfk )C_{\psi}C_J T_c$ by comparing with $C_0^{-1}h_c T_c$  .
Then, we can factorize the determinant as 
\begin{eqnarray}
\det {\cal M} &&\sim (-i\omega+\lambda_0(\bfk)\chi_0^{-1}(\bfk)) \nonumber \\
                         &&\times (-i\omega+i k c_s+\frac{1}{2}\nu_0^{l}h_c^{-1} k^2) \nonumber \\
                          && \times             (-i\omega-i k c_s+\frac{1}{2}\nu_0^{l}h_c^{-1} k^2),
\end{eqnarray}
in the long-wavelength region.
Here, we have defined
\begin{eqnarray}
\lambda_0(\bfk)&& \equiv\lambda_0 k^2 D_{\psi}, \\
c_s^2 &&\equiv C_0^{-1}h_c T_c.
\end{eqnarray}
The diagonal components of the cofactor matrix $m$ reads
\begin{eqnarray}
m_{11}\sim &&(-i\omega+i k c_s+\frac{1}{2}\nu_0^l h_c^{-1} k^2)  \nonumber \\
               &&\times (-i\omega-i k c_s+\frac{1}{2}\nu_0^l h_c^{-1} k^2), \\
m_{22}=&&(-i\omega)^2-i\omega k^2\lambda_0\chi_0(\bfk)D_{\psi}\nu_0^l h_c^{-1} \nonumber \\
            &&+k^2\chi_0^{-1}(\bfk)C_{\psi}C_{J}T_c \nonumber \\
            &&+k^4\lambda_0\chi_0(\bfk)D_{\psi}\nu_0^l h_c^{-1} , \\
m_{33}=&&(-i\omega)(-i\omega+\lambda_0(\bfk)\chi_0^{-1}(\bfk)),
\end{eqnarray}
and the off-diagonal components are given by
\begin{eqnarray}
m_{12}&&=k^2\chi_0^{-1}(\bfk)\beta_2^{-1}C_J T_c,\\
m_{13}&&=-k\omega\chi_0^{-1}(\bfk)C_J T_c,\\
m_{21}&&=-k^2 C_0^{-1}h_c C_{\psi}\beta_2 T_c,\\
m_{23}&&=-i k C_0^{-1}h_c\beta_2 T_c(-i\omega+k^2\lambda_0\chi_0(\bfk)D_{\psi}),\\
m_{31}&&=k^2 C_0^{-1}h_c C_\psi\beta_2 T_c,\\
m_{32}&&=-i k\beta_2^{-1}(-i\omega+\lambda_0 (\bfk )\chi_0^{-1}(\bfk )).
\end{eqnarray}
Here, we neglect the off-diagonal components 
because they would not yield dominant contributions to the transport coefficients.
Then, we obtain the bare propagator as
\begin{equation}
G^0(\bfk, \omega)= 
  \begin{pmatrix}
  G^0_{\psi}(\bfk,\omega) & 0 & 0 \\
  0 & G^0_{m}(\bfk,\omega) & 0\\
  0 & 0 & G^0_{\parallel}(\bfk,\omega)
  \end{pmatrix}.\label{eq: g0}
\end{equation}
with
\begin{eqnarray}
G^0_{\psi}(\bfk,\omega)  =&&\frac{1}{-i\omega+\lambda(\bfk)\chi_0^{-1}(\bfk)}, 
\label{eq: gthermal}\\
G^0_{\parallel}(\bfk,\omega) 
  \sim &&\frac{1}{2}\bl[ \frac{1}{-i\omega+i k c_s+\frac{1}{2}\nu_0^l h_c^{-1}k^2} \nonumber \\
                         &&+\frac{1}{-i\omega-i k c_s+\frac{1}{2}\nu_0^{l}h_c^{-1} k^2}\br].
\label{eq: gsound}
\end{eqnarray}
$G^0_{mm}(\bfk,\omega)$ is not needed in later calculations.
The bare propagator of $\delta \bfJ_{\perp}$ is trivially given by 
\begin{equation}
G^0_{\perp}(\bfk,\omega)=\frac{1}{-i\omega+\eta_0 k^2 h_c^{-1}}.
\label{eq: gviscous}
\end{equation}
We finally arrive at the equations of motion as the self-consistent form:
\begin{equation}
  \begin{pmatrix}
  \tilde{\psi}(\bfk ,\omega) \\
  \tilde{m}(\bfk ,\omega) \\
  \delta \tilde{J}_{\parallel}(\bfk,\omega)
  \end{pmatrix}
=
 \begin{pmatrix}
  \tilde{\psi}^0(\bfk ,\omega) \\
  0 \\
  \delta \tilde{J}_{\parallel}^0(\bfk,\omega)
  \end{pmatrix}
+G^0(\bfk,\omega)\bfV(\bfk,\omega),
\label{eq: selfeom}
\end{equation}
and
\begin{equation}
\delta \tilde{\bfJ}_{\perp}(\bfk,\omega)
=\delta \tilde{\bfJ}_{\perp}^0(\bfk,\omega)+G^0_{\perp}\bfV_{\perp \psi\psi}(\bfk,\omega),
\label{eq: selfjeom}
\end{equation}
where
\begin{eqnarray}
\tilde{\psi}^0(\bfk ,\omega) &&=G^0_{\psi}(\bfk,\omega)  \alpha_1^{-1} \tilde{\theta}_n(\bfk,\omega),
\label{eq: a}\\
\delta \tilde{J}_{\parallel}^0(\bfk,\omega) &&=G^0_{\parallel}(\bfk,\omega) \tilde{\theta}_{\parallel}(\bfk,\omega),\\
\delta \tilde{\bfJ}_{\perp}^0(\bfk,\omega) &&=G^0_{\perp}(\bfk,\omega)\tilde{\bftheta}_{\perp}(\bfk,\omega), \\
\bfV_{\perp \psi\psi}(\bfk,\omega)
=&&-i T_c{\cal P}_{\perp}(\bfk)\cdot\int_{q\Omega}\bfq \chi_0^{-1}(\bfq) \nonumber \\
   &&\times \tilde{\psi}(q)\tilde{\psi}(k-q). \label{eq: v4} 
\end{eqnarray}
Here, $\tilde{\psi}^0(\bfk ,\omega) $, $\delta \tilde{J}_{\parallel}^0(\bfk,\omega) $
and $\delta \tilde{\bfJ}_{\perp}^0 $ are the bare variables that are the solutions without the nonlinear terms.
Iterating Eqs. (\ref{eq: selfeom}) and (\ref{eq: selfjeom}),
we can obtain perturbative expansions about nonlinear interactions $\bfV$ and $\bfV_{\perp \psi\psi}$.
We note that the first and third rows of Eq. (\ref{eq: selfeom}) 
are the equations of motion for the thermal and sound modes, respectively, 
while Eq. (\ref{eq: selfjeom}) is for the viscous mode.
We also stress that Eqs. (\ref{eq: gthermal})-(\ref{eq: gviscous}) are the propagators of
the thermal, sound and viscous modes, respectively.    

Now, we calculate the two body correlation of 
$\tilde{\psi}^0(\bfk ,\omega)$ and $\delta \tilde{\bfJ}_{\perp}^0(\bfk,\omega) $
which are needed in later calculations.
\begin{eqnarray}
\la\tilde{\psi}^0(\bfk_1 ,\omega_1)\tilde{\psi}^0(\bfk_2 ,\omega_2) \ra
  =&&G^0_{\psi}(\bfk_1 ,\omega_1)G^0_{\psi}(\bfk_2 ,\omega_2)\alpha_1^{-2} \nonumber \\
    &&\times \la\tilde{\theta}(\bfk_1,\omega_1)\tilde{\theta}(\bfk_2,\omega_2) \ra.
\end{eqnarray}
Using the fluctuation dissipation relation Eq.(\ref{eq: fdrn}), we find
\begin{equation}
\la\tilde{\theta}(\bfk_1,\omega_1)\tilde{\theta}(\bfk_2,\omega_2) \ra
=2  \alpha_1^2 \lambda_0(\bfk)(2\pi)^{d+1}\delta( k_1+k_2),
\end{equation}
and
\begin{eqnarray}
\la\tilde{\psi}^0(\bfk_1 ,\omega_1)\tilde{\psi}^0(\bfk_2 ,\omega_2) \ra
=&&\frac{2  \lambda_0(\bfk_1)}{\omega_1^2+\lambda_0^2(\bfk_1)\chi_0^{-2}(\bfk_1)} \nonumber \\
&&\hspace{-0.1cm}\times (2\pi)^{d+1}\delta( k_1+k_2),
\label{eq: pcorrelation}
\end{eqnarray}
where $\delta( k_1+k_2) \equiv \delta( \bfk_1+\bfk_2)\delta( \omega_1+\omega_2)$.
By a similar calculation, we obtain
\begin{eqnarray}
\la \delta\tilde{J}_{\perp}^{i}(\bfk_1,\omega_1) \delta\tilde{J}_{\perp}^{i}(\bfk_2,\omega_2) \ra
&&=\frac{2 T_c \eta_0(\bfk_1)}{\omega_1^2+\eta_0^2(\bfk_1)h_c^{-2}}({\cal P}_{\perp}(\bfk_1))_{i j} \nonumber \\
&&\times (2\pi)^{d+1}\delta( k_1+k_2),
\end{eqnarray}
where $\eta_0(\bfk)=\eta_0 k^2$.
For a later convenience,
we define 
\begin{eqnarray}
C^0_{\psi}(\bfk, \omega ) = &&\frac{2  \lambda_0(\bfk )}{\omega^2+\lambda_0^2(\bfk )\chi_0^{-2}(\bfk )}, \\
C^0_{\perp}(\bfk, \omega )= &&\frac{2 T_c \eta_0(\bfk )}{\omega^2+\eta_0^2(\bfk )h_c^{-2}}.
\end{eqnarray}

\section{Renormalization of the thermal and viscous diffusion modes}
\label{sec: RG1}
\begin{figure*}[!t]
    \begin{minipage}{1.0\hsize}
     \centering
     \includegraphics[width=\hsize]{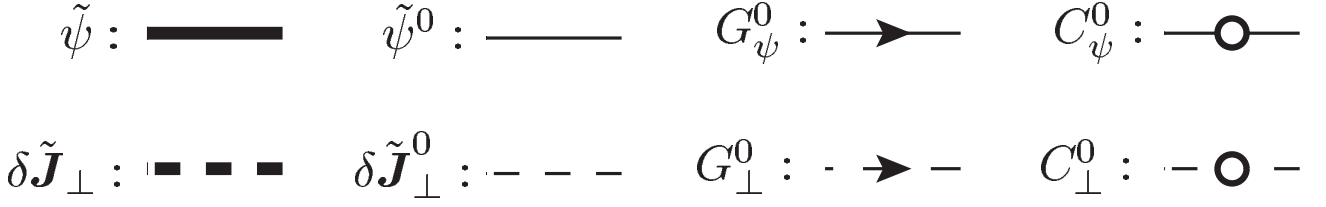}
       \caption{Diagrams for the full and bare variables, 
                   the bare propagators and the bare correlations.}
     \label{fig: difinition}
    \end{minipage}
\end{figure*}

\begin{figure}[!t]
    \begin{minipage}{1.0\hsize}
     \centering
     \includegraphics[width=\hsize]{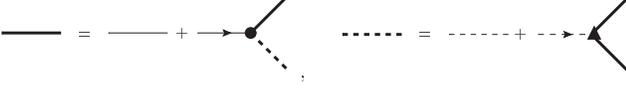}
       \caption{Diagrams of the equations of motion for the thermal and viscous modes.
                   The left and right hand side respectively denote Eqs. (\ref{eq: selfpeom}) and (\ref{eq: selfjeom}). }
     \label{fig: eom}
    \end{minipage}
\end{figure}

Here, we first derive the RG equations for the thermal conductivity and the shear viscosity.
Now, we note that the sound mode is not a genuine-relevant mode but a secondly mode 
that is strongly affected by order-parameter fluctuations 
but yields only a negligible feedback for the order parameters \cite{Minami:2009hn, kroll}.
Then, we can neglect the sound mode for the minimal critical dynamics;
however, the bulk viscosity is not renormalized in that case.
Here, to first analyze the minimal  dynamics, 
we neglect the secondly mode, which is renormalized in the next section. 
In that case, the equations of motion are given by
\begin{eqnarray}
\tilde{\psi}(\bfk, \omega) =&&\tilde{\psi}^{0}(\bfk, \omega)+G_{\psi}^{0}(\bfk, \omega )V_{\psi\psi\perp}(\bfk, \omega )
\label{eq: selfpeom}
\end{eqnarray}
and Eq. (\ref{eq: selfjeom}).
For a diagrammatic treatment ,
we denote the full and bare variables, the bare propagators 
and the bare correlation functions as Fig. \ref{fig: difinition}. 
Then, we can represent the equations of motion (\ref{eq: selfpeom}) and (\ref{eq: selfjeom}) as Fig. \ref{fig: eom}.

\begin{figure*}[!t]
    \begin{minipage}{1.0\hsize}
     \centering
     \includegraphics[width=\hsize]{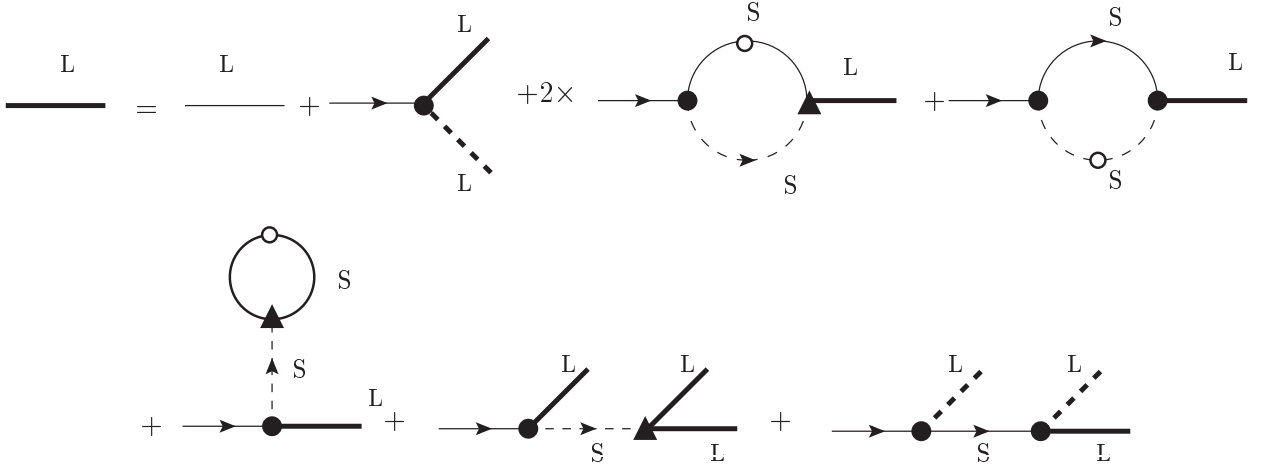}
       \caption{Diagrams for the coarse-grained equation of motion for $\psi$.
                   The letters, L and S, respectively denote the long- and short-wavelength components
                   (see the text below Eq.(\ref{eq: abcde})).}
     \label{fig: cgpeom}
    \end{minipage}
\end{figure*}

For  coarse gaining, we decompose the variables into the long- 
and short-wavelength components as
\begin{equation}
\tilde{\psi}(\bfk, \omega) = \tilde{\psi}^{\rm L} (\bfk, \omega )+\tilde{\psi}^{\rm S}(\bfk, \omega), 
\end{equation}
with
\begin{eqnarray}
\tilde{\psi}^{\rm L}(\bfk, \omega ) \equiv&& \Theta(\Lambda-\delta\Lambda-k)\tilde{\psi}(\bfk, \omega),\\  
\tilde{\psi}^{\rm S} (\bfk, \omega)\equiv&& \Theta(k-\Lambda-\delta\Lambda)\tilde{\psi}(\bfk, \omega) ,
\label{eq: abcde}
\end{eqnarray}
where $\Theta(x)$ is a step function; 
i.e., the wavenumber is decomposed into $0 < k < \Lambda - \delta \Lambda$ 
and $\Lambda - \delta \Lambda < k < \Lambda$. 
Hereafter, quantities with the suffixes L and S are supposed to be decomposed as above.
To average over the $\tilde{\psi}^{0 {\rm S}}$ and $\delta \tilde{\bfJ}_{\perp}^{0 {\rm S}}$,
we must solve the equation of motion about them.  
Here, we solve the equations of motion to second order in the nonlinear interactions
and average over $\tilde{\psi}^{0 {\rm S}} $ and $\delta \tilde{\bfJ}_{\perp}^{0 {\rm S}}$.
Then, we find the coarse-grained equation of motion for $ \psi$, 
which is diagrammatically given by Fig. \ref{fig: cgpeom}.
The last two terms in Fig. \ref{fig: cgpeom} represent nonlinear interactions being of third order, 
and can be neglected.
Furthermore, the fifth term vanishes due to the relation between the step and delta functions in the loop integral.
\begin{figure*}[!t]
    \begin{minipage}{1.0\hsize}
     \centering
     \includegraphics[width=\hsize]{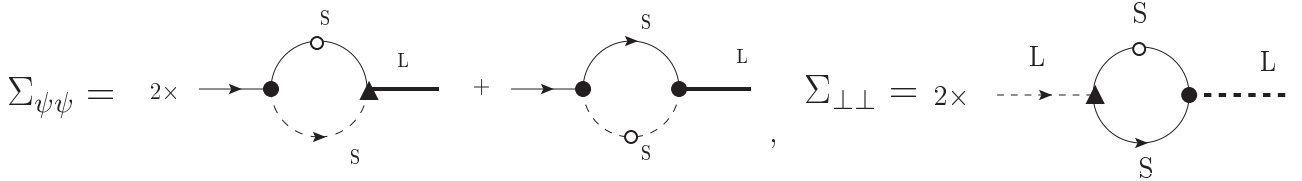}
       \caption{Diagrams for the self energies.}
     \label{fig: self}
    \end{minipage}
\end{figure*}% 
Introducing the self energy $\Sigma_{\psi\psi}$, which is graphically represented in Fig. \ref{fig: self},
we can write the coarse-grained equation of motion for $ \psi$ as
\begin{eqnarray}
\tilde{\psi}^{\rm L}(\bfk, \omega) 
  =&&\tilde{\psi}^{0{\rm L}}(\bfk, \omega)+G_{\psi}^{0{\rm L}}(\bfk, \omega )V_{\psi\psi\perp}^{\rm L} (\bfk, \omega )
\nonumber \\
   &&+\tilde{\psi}^{\rm L}(\bfk, \omega)G_{\psi}^{0{\rm L}}(\bfk, \omega )\Sigma_{\psi\psi}(\bfk, \omega ).
   \label{eq: cgpeom}
\end{eqnarray} 
The self energy is given by 
\begin{eqnarray}
\Sigma_{\psi\psi}&&(\bfk, \omega ) =-T_c h_c^{-1} k^2 \chi_0^{-1} (\bfk)  \nonumber \\
  &&\hspace{-0.3cm}\times\int_q  \frac{(\hat{\bfk}\cdot{\cal P}(\bfk-\bfq )\cdot\hat{\bfk})\chi_0(\bfq)}
  {-i\omega+\lambda_0(\bfq)\chi_0^{-1}(\bfq)+\eta_0(\bfk-\bfq)h_c^{-1} },
\end{eqnarray}
where $\eta_0(\bfk)=\eta_0 k^2$.
Solving Eq. (\ref{eq: cgpeom}) about $\tilde{\psi}^{\rm L}$,  we have
\begin{eqnarray}
\tilde{\psi}^{\rm L}=&&[(G_{\psi}^{0{\rm L}})^{-1} -\Sigma_{\psi\psi}]^{-1}\alpha_1^{-1}\tilde{\theta}_n
 \nonumber \\ &&
+[(G_{\psi}^{0{\rm L}})^{-1} -\Sigma_{\psi\psi}]^{-1}V_{\psi\psi\perp}^{\rm L}.
\end{eqnarray}
where we have used Eq. (\ref{eq: a}).
Introducing renormalized variables as
\begin{eqnarray}
(G_{\psi {\rm R}})^{-1}(\bfk, \omega )
  =&&(G_{\psi}^{0 {\rm L} })^{-1}(\bfk, \omega ) -\Sigma_{\psi \psi}(\bfk, \omega ), \label{eq: rppro}\\
\tilde{\psi}^{0{\rm L}}_ {\rm R}(\bfk, \omega) =&&G_{\psi{\rm R}}(\bfk, \omega )\alpha_1^{-1}\tilde{\theta}_n(\bfk, \omega ), 
\end{eqnarray}
we can rewrite Eq.(\ref{eq: a}) as the renormalized equation of motion:
\begin{eqnarray}
\tilde{\psi}^{\rm L}=&&\tilde{\psi}^{0{\rm L}}_ {\rm R}(\bfk, \omega )
+G_{\psi {\rm R}}(\bfk, \omega )V_{\psi\psi\perp}^{\rm L}.
\label{eq: bbb}
\end{eqnarray}
We now require that the renormalized propagator has the same form as the bare one:
\begin{equation}
(G_{\psi {\rm R}})^{-1}(\bfk, \omega )= -i\omega +\lambda_R D_\psi k^2 \chi_0^{-1}(\bfk ),
\end{equation} 
where $\lambda_{\rm R}$ is the renormalized thermal conductivity.
That is, we require that the only transport coefficients are explicitly renormalized.
The small correction for the thermal conductivity $\delta \lambda \equiv \lambda_{\rm R}-\lambda_0$ reads
\begin{eqnarray}
\delta \lambda =&&-\lim_{k, \omega \to 0}[(D_\psi k^2 \chi_0(\bfk ))^{-1}\Sigma_{\psi \psi}(\bfk, \omega )],
                                                                                                                            \nonumber \\
                    =&&\frac{T_c}{ h_c D_\psi }\int_q 
                         \frac{(\hat{\bfk}\cdot{\cal P}(\bfq )\cdot\hat{\bfk})\chi_0(\bfq)}
                           {\lambda_0(\bfq)\chi_0^{-1}(\bfq)+\eta_0(\bfq)h_c^{-1} }.        
\end{eqnarray}
We  approximate the denominator and the numerator as
\begin{eqnarray}
\lambda_0(\bfq)\chi_0^{-1}(\bfq)&&+\eta_0(\bfq)h_c^{-1} \sim  \eta_0 (\bfk) h_c^{-1}, \\
\chi_0^{-1}(\bfq ) = && r_0 + q^2 \sim q^2, 
\end{eqnarray}
near the CP \cite{siggia}.
Then, we find
\begin{eqnarray}
\delta \lambda &&\sim  \frac{T_c}{D_\psi \eta_0} 
     \int \frac{d\Omega_d}{(2\pi)^d}(\hat{\bfk}\cdot{\cal P}(\bfq )\cdot\hat{\bfk})
     \int^{\Lambda}_{\Lambda-\delta \Lambda} d q q^{d-5}   \nonumber \\
  &&=- \frac{T_c}{D_\psi \eta_0} \int \frac{d\Omega_d}{(2\pi)^d}(\hat{\bfk}\cdot{\cal P}(\bfq )\cdot\hat{\bfk}) 
     \Lambda^{d-5} \delta \Lambda,
\end{eqnarray}
where $d\Omega_d$ is the solid angle in the space dimension $d$.
Therefore, we obtain the RG equation for the thermal conductivity:
\begin{equation}
-\Lambda \frac{\partial \lambda}{\partial \Lambda}=\frac{T_c}{D_\psi \eta(\Lambda)} \int \frac{d\Omega_d}{(2\pi)^d}(\hat{\bfk}\cdot{\cal P}(\bfq )\cdot\hat{\bfk}) 
     \Lambda^{d-4}, 
\end{equation}
where $\eta_0$ is rewritten as $\eta (\Lambda)$. 
For the space dimensions, $d=4-\epsilon$,
the angle integral is given by 
\begin{equation}
 \int \frac{d\Omega_4}{(2\pi)^4}(\hat{\bfk}\cdot{\cal P}(\bfq )\cdot\hat{\bfk})  =\frac{3}{4}K_4,
\end{equation}
where $K_4$ is the surface area of a unit sphere in 4 dimensions divided by $(2\pi)^4$.
The RG equation in $4- \epsilon$ dimensions reads
\begin{equation}
-\Lambda \frac{\delta \lambda}{\delta \Lambda}=\frac{3}{4} f(\Lambda) \lambda(\Lambda), 
\label{eq: rgl2}
\end{equation}
where we have introduced
\begin{equation}
f(\Lambda) \equiv \frac{T_c K_4}{D_\psi \eta(\Lambda) \lambda(\Lambda) \Lambda^{\epsilon}},
\label{eq: flambda}
\end{equation} 
for a later convenience.

By making coarse graining of the viscous mode with a similar procedures as above,
we obtain a small correction for the shear viscosity:
\begin{equation} 
\delta \eta = -\lim_{k, \omega \to 0}[(k^2 h_c^{-1}(d-1))^{-1}\sum_i(\Sigma_{\perp\perp}(\bfk, \omega))_{ii}],
\label{eq: deltaeta}
\end{equation} 
where $(\Sigma_{\perp\perp}(\bfk, \omega))_{ij}$ is the self energy for the viscous mode and given by
\begin{eqnarray}
(\Sigma_{\perp\perp}&&(\bfk, \omega))_{i j}=-T_c h_c^{-1}
\int_q \chi_0(\bfk-\bfq) ({\cal P}_{\perp}(\bfk )\cdot \bfq )_i q_j \nonumber \\
&& \hspace{-1.0cm} \times\frac{\chi_0^{-1}(\bfq) -\chi_0^{-1}(\bfk-\bfq)}
{-i\omega +\lambda_0(\bfq )\chi_0^{-1}(\bfq )+\lambda(\bfk-\bfq)\chi_0^{-1}(\bfk-\bfq )},
\label{eq: jself}
\end{eqnarray}
which is graphically represented as Fig. \ref{fig: self} .
In the space dimension $d=4-\epsilon$,
we find the RG equation for the shear viscosity
\begin{equation}
-\Lambda \frac{\partial \eta (\Lambda)}{\partial \Lambda}=\frac{1}{24} f(\Lambda) \eta(\Lambda),
\label{eq: rge2}
\end{equation}
where the prefactor $1/24$ comes from the angular integral in Eq. (\ref{eq: jself})
 and the factor $(d-1)^{-1}$ in Eq. (\ref{eq: deltaeta}).
 
Differentiating Eq. (\ref{eq: flambda}) about $\Lambda$,
we have the RG equation for $f(\Lambda)$
\begin{equation}
-\Lambda \frac{\partial f(\Lambda)}{\partial \Lambda}=(\epsilon-\frac{19}{24}f(\Lambda ))f(\Lambda ).
\end{equation} 
 
\section{Renormalization of the sound mode}
\label{sec: RG2}
Next, let us make a coarse graining of the sound mode for the renormalized bulk viscosity.   
Because a feedback from the sound mode is neglected, 
we must renormalize the mode with a method separating 
relevant and secondly modes \cite{kroll}.
Here, we take the method developed by Onuki \cite{onukin, onuki} 
, in which RG equations are derived from fluctuation-dissipation relations.

Now, we consider the equation of motion for the sound mode, (\ref{eq: soundeom}):
\begin{eqnarray}
-i\omega\delta\tilde{J}_{\parallel}
   =&&- i k T_c[\chi_0^{-1}(\bfk)C_J \tilde{\psi}+ C_0^{-1}\beta_2 h_c\tilde{m} \nonumber \\
     &&+\beta_2^{-1}h_c\gamma_0 \int_{q\Omega}\tilde{\psi}(q)\tilde{\psi}(k-q) ] \nonumber \\
     &&-k^2\nu_0^l h_c^{-1}\delta\tilde{J}_{\parallel} +\tilde{\theta}^{0}_{\parallel}, 
     \label{eq: soundeom2}
\end{eqnarray}
where the noise term $\tilde{\theta}_{\parallel}^0$ satisfies the fluctuation dissipation relation:
\begin{eqnarray}
\la \tilde{\theta}^{0}_{\parallel}(\bfk_1, \omega_1) \tilde{\theta}^{0}_{\parallel}(\bfk_2, \omega_2) \ra &&
=2 T_c k_1^2 \nu_0^{l} \nonumber \\ &&
\times (2\pi)^{d+1}\delta(k_1+k_2).
\end{eqnarray}
Since $\delta \bfJ_{\parallel}$ is a conserved density projected onto $\hat{\bfk}$,
we can rewrite  Eq. (\ref{eq: soundeom2}) as 
\begin{equation}
-i\omega \delta\tilde{J}_{\parallel} (\bfk, \omega) =i \bfk \cdot \tilde{\Pi}(\bfk, \omega )\cdot \hat{\bfk},
\end{equation}
where $\tilde{\Pi}_{i j}$ is the stress tensor.
If we take  $z$ direction as  $\hat{\bfk}$,
$\tilde{\Pi}_{z z}$ reads
\begin{eqnarray}
\tilde{\Pi}_{z z}(\bfk, \omega ) &&= -T_c[\chi_0^{-1}(\bfk)C_J \tilde{\psi}(\bfk, \omega)
                                            + C_0^{-1}\beta_2 h_c\tilde{m} (\bfk, \omega)\nonumber \\
     &&+\beta_2^{-1}h_c\gamma_0 \int_{q\Omega}\tilde{\psi}(q)\tilde{\psi}(k-q) ] \nonumber \\
     &&+i k\nu_0^l h_c^{-1}\delta\tilde{J}_{\parallel}(\bfk, \omega) +\tilde{\pi}^{0}_{z z}(\bfk, \omega) ,
\end{eqnarray} 
where $\tilde{\pi}^{0}_{i j}(\bfk, \omega)$ is the random-stress tensor coming from microscopic process
and satisfies the relation, 
$i \bfk \cdot \tilde{\pi}^{0}(\bfk, \omega) \cdot \hat{\bfk}=\tilde{\theta}^{0}_{\parallel}(\bfk, \omega)$.

We now consider how Eq. (\ref{eq: soundeom2}) is affected by the coarse-graining procedure.
In the coarse-graining procedure,
 the variables, $\tilde{\psi}^{\rm S}$, $\tilde{m}^{\rm S}$ and $\delta \tilde{\bfJ}^{\rm S}$ 
are eliminated from Eq. (\ref{eq: soundeom2}).
The eliminated variables do not disappear from the equation of motion but are
implicitly contained in the noise term.
In other words, we convert the macroscopic process in the wavenumber shell
$\Lambda - \delta \Lambda < k < \Lambda$ into the microscopic process.
In this procedure, the noise term is implicitly renormalized as follows
\begin{equation}
\tilde{\theta}^{R}_{\parallel}(\bfk, \omega)=\tilde{\theta}^{0}_{\parallel}(\bfk, \omega)
+\tilde{\theta}^{\rm Macro}_{\parallel}(\bfk, \omega) \label{eq: macro},
\end{equation}
where
\begin{eqnarray}
\tilde{\theta}^{\rm Macro}_{\parallel}(\bfk, \omega )
   &&\equiv  i\bfk \cdot \tilde{\pi}^{\rm Macro}(\bfk, \omega ) \cdot \hat{\bfk}, \\
\tilde{\pi}^{\rm Macro}_{z z}(\bfk, \omega )&&\equiv -T_c[\chi_0^{-1}(\bfk)C_J \tilde{\psi}^{\rm S}
                                            + C_0^{-1}\beta_2 h_c\tilde{m}^{\rm S} \nonumber \\
     &&+\beta_2^{-1}h_c\gamma_0 \int_{q\Omega}\tilde{\psi}^{\rm S}(q)\tilde{\psi}^{\rm S}(k-q) ] \nonumber \\
     &&+i k\nu_0^l h_c^{-1}\delta\tilde{J}_{\parallel}^{\rm S}, \label{eq: pi}\\
     \sim && -T_c \beta_2^{-1}h_c\gamma_0 \int_{q\Omega}\tilde{\psi}^{\rm S}(q)\tilde{\psi}^{\rm S}(k-q),
\end{eqnarray}
where we neglect the linear terms in Eq. (\ref{eq: pi}) that is irrelevant for the following argument.
The new term $\tilde{\theta}^{\rm Macro}_{\parallel}(\bfk, \omega )$, being due to the coarse graining,
contributes the transport coefficient through the fluctuation-dissipation relation:
\begin{eqnarray}
\la &&\tilde{\theta}^{\rm Macro}_{\parallel}(\bfk_1, \omega_1) 
   \tilde{\theta}^{\rm Macro}_{\parallel}(\bfk_2, \omega_2) \ra \nonumber \\ 
         &&=2 T_c k_1^2 \delta\nu^{l}(\bfk_1, \omega_1 )(2\pi)^{d+1}\delta(k_1+k_2),
         \label{eq: bb}
\end{eqnarray}
where we have assumed that the renormalized equation of motion has the same form as Eq. (\ref{eq: soundeom2}).
We note that this assumption is equivalent to the requirement below Eq. (\ref{eq: bbb}).
Now, we calculate the left-hand side in Eq. (\ref{eq: bb}):
\begin{eqnarray}
\la \tilde{\theta}^{\rm Macro}_{\parallel}(\bfk_1, \omega_1) 
\tilde{\theta}^{\rm Macro}_{\parallel}(\bfk_2, \omega_2) \ra
=&&-k_1 k_2 (T_c h_c \beta_2^{-1})^2\gamma_0^{2} \nonumber \\
&&\hspace{-4.5cm}
\times  \int_{q_1 \Omega_1 q_2 \Omega_2} \hspace{-1.0cm}
\la \tilde{\psi}^{\rm S}(q_1) \tilde{\psi}^{\rm S}(k_1-q_1)
\tilde{\psi}^{\rm S}(q_2) \tilde{\psi}^{\rm S}(k_2-q_2) \ra.
\end{eqnarray}
Approximating the variable by the bare one, $\tilde{\psi}^{\rm S} \sim \tilde{\psi}^{0 {\rm S}}$,
we find
\begin{eqnarray}
\la && \tilde{\theta}^{\rm Macro}_{\parallel}(\bfk_1, \omega_1) 
     \tilde{\theta}^{\rm Macro}_{\parallel}(\bfk_2, \omega_2) \ra
     = (2\pi)^{d+1}\delta( k_1+k_2) \nonumber \\
  && \times 2 k_1^2 (T_c h_c \beta_2^{-1})^2\gamma_0^{2} 
     \int_{q \Omega}C_{\psi}^{0{\rm S}}(q) C_{\psi}^{0{\rm S}}(k_1-q),
\end{eqnarray}
where we have used Eq. (\ref{eq: pcorrelation}) and neglected a term corresponding to a disconnected diagram.
Then, comparing with Eq. (\ref{eq: bb}), we obtain the correction to the longitudinal-kinetic viscosity:
\begin{equation}
\delta \nu^{l}(\bfk, \omega ) =T_c \beta_2^{-2} h_c^2 \gamma_0^2
\int_{q \Omega}  C_{\psi}^{0{\rm S}}(q) C_{\psi}^{0{\rm S}}(k-q).
\end{equation}
We are not interested in the frequency- or wavenumber-dependent bulk viscosity
and then take the limit $k, \omega \rightarrow 0$:
\begin{eqnarray}
\delta \nu^{l} &&\equiv \lim_{k, \omega \to 0}\delta \nu^{l}(\bfk, \omega ) \nonumber \\
 &&=T_c \beta_2^{-2} h_c^2 \gamma_0^2
\int_{q \Omega}  (C_{\psi}^{0{\rm S}}(q))^2 .
\end{eqnarray}
After the integration, we find the RG equation for longitudinal kinetic viscosity:
\begin{eqnarray}
-\Lambda\frac{\partial \nu^l(\Lambda)}{\partial \Lambda}
=\frac{T_c h_c^2 K_4}{\beta_2^2 D_\psi}\gamma^2(\Lambda)\lambda^{-1}(\Lambda)\Lambda^{-\epsilon-4}.
\end{eqnarray} 
where we have rewritten the static parameter $\gamma_0$ as $\gamma (\Lambda)$ to 
denote its cutoff dependence as mentioned in the text.
The asymptotic behavior obtained from this RG equation is different from 
the shear viscosity's behavior, so we  replace above RG equation as 
 \begin{eqnarray}
-\Lambda\frac{\partial \zeta(\Lambda)}{\partial \Lambda}
=\frac{T_c h_c^2 K_4}{\beta_2^2 D_\psi}\gamma^2(\Lambda)\lambda^{-1}(\Lambda)\Lambda^{-\epsilon-4}.
\end{eqnarray} 

Although, by this method,
we could more easily obtain the RG equations for the thermal conductivity and shear viscosity,
we have taken the diagrammatic method for an instructive purpose.

\end{document}